\documentclass[conference]{IEEEtran}
\IEEEoverridecommandlockouts
\pdfoutput=1
\usepackage{cite}
\usepackage{amsmath,amssymb,amsfonts}
\usepackage{graphicx}
\usepackage{textcomp}
\usepackage{subfigure}
\usepackage[table,xcdraw]{xcolor}
\usepackage[]{algorithm2e}
\def\BibTeX{{\rm B\kern-.05em{\sc i\kern-.025em b}\kern-.08em
    T\kern-.1667em\lower.7ex\hbox{E}\kern-.125emX}}
\usepackage{multirow}
\usepackage{colortbl}
\usepackage{hhline}
\usepackage{tikz}
\usepackage{amssymb}
\usepackage{pifont}
\usepackage{diagbox}
\usepackage{url}

\newcommand{\Mycaption}[1]{
  \vspace{-2.0ex}
  \caption{#1}
  \vspace{-2.0ex}
}
\usepackage{pifont}

\begin{document}

%\title{MPI Meets Python: Enhancing OMB with Python Benchmarks}
\title{OMB-Py: Python Micro-Benchmarks for Evaluating Performance of MPI Libraries on HPC Systems\\
\thanks{*This research is supported in part by NSF grants \#1818253, \#1854828, \#1931537, \#2007991, \#2018627, \#2112606, and XRAC grant \#NCR-130002.}
}

\author{
  \IEEEauthorblockN{
    Nawras Alnaasan,
    Arpan Jain,
    Aamir Shafi,
    Hari Subramoni,
    and Dhabaleswar K Panda\\
  }%
  %\and
  \IEEEauthorblockA{
    \small{
      \textit{Department of Computer Science and Engineering},}
      \\
      \small{
      The Ohio State University, Columbus, Ohio, USA,}
      \\
      \small{
      \texttt{\{alnaasan.1, jain.575, shafi.16, subramoni.1\}@osu.edu, panda@cse.ohio-state.edu}
    }%
  }%

}

\maketitle
\pagestyle{plain}
\begin{abstract}
Python has become a dominant programming language for emerging areas like Machine Learning (ML), Deep Learning (DL), and Data Science (DS). An attractive feature of Python is that it provides easy-to-use programming interface while allowing library developers to enhance performance of their applications by harnessing the computing power offered by High Performance Computing (HPC) platforms. Efficient communication is key to scaling applications on parallel systems, which is typically enabled by the Message Passing Interface (MPI) standard and compliant libraries on HPC hardware. mpi4py is a Python-based communication library that provides an MPI-like interface for Python applications allowing application developers to utilize parallel processing elements including GPUs. However, there is currently no benchmark suite to evaluate communication performance of mpi4py---and Python MPI codes in general---on modern HPC systems. In order to bridge this gap, we propose OMB-Py---Python extensions to the open-source OSU Micro-Benchmark (OMB) suite---aimed to evaluate communication performance of MPI-based parallel applications in Python. To the best of our knowledge, OMB-Py is the first communication benchmark suite for parallel Python applications. OMB-Py consists of a variety of point-to-point and collective communication benchmark tests that are implemented for a range of popular Python libraries including NumPy, CuPy, Numba, and PyCUDA. Our evaluation reveals that mpi4py introduces a small overhead when compared to native MPI libraries. We plan to publicly release OMB-Py to benefit the Python HPC community.
\end{abstract}

\begin{IEEEkeywords}
MPI, Python, mpi4py, OMB, Benchmarks, HPC
\end{IEEEkeywords}

\section{Introduction}
\label{sec:intro}

\subsection{Motivation}
Message Passing Interface (MPI) is considered as the de facto standard that defines communication operations for exchanging data in parallel computing environments. The MPI standard~\cite{MPI} defines multiple operations for various communication purposes and provides bindings for C and Fortran programming languages. However, other languages like Python are dominant when it comes to application areas like Machine Learning (ML), Deep Learning (DL), and Data Science (DS) with compute-intensive tasks that requires optimized communication for scalability. To use MPI with higher-level programming languages like Python, a communication wrapper library is needed to provide the necessary MPI-like bindings. A popular representative Python MPI-like library is the mpi4py~\cite{mpi4py} package that has been used by a range of applications and projects including HDF5 for Python~\cite{hdf54py}, Dask~\cite{dask}, mpi4py-fft~\cite{fft,fft_paper}, yt~\cite{yt}, and the Visualization Toolkit (VTK)~\cite{vtk}.

Computationally intensive tasks are becoming more demanding every day; therefore, any gain in performance is highly beneficial. To aid in maximizing efficiency, developers must have the necessary tools to characterize the performance of the systems they are working on. Micro-benchmarks play an important role in understanding the behavior of critical points in a system. Due to the myriad of interconnected components in an HPC system, micro-benchmarks are instrumental to gain insight into the impact of each of these components. In addition, MPI implementations offer a wide range of parameters to achieve optimized performance depending on the system, the underlying interconnects, the devices in communication, and the executed applications. Micro-benchmarks packages such as OSU Micro Benchmarks (OMB)~\cite{OMB} can be used to evaluate the performance of MPI implementations on HPC systems. mpi4py has been serving the Python HPC community for well over a decade, yet there is no benchmark suite to evaluate communication performance of mpi4py---and Python MPI codes in general---on modern HPC systems.

\subsection{Contributions}
The mpi4py software currently only provides some basic sample applications for MPI operations. In order to bridge this gap, we propose OMB-Py---Python extensions to the open-source OMB suite---aimed to evaluate communication performance of MPI-based parallel applications in Python. To the best of our knowledge, OMB-Py is the first communication benchmark suite for parallel Python applications. OMB-Py is aimed to evaluate performance of MPI using Python on both CPUs and GPUs. Table~\ref{table:benchmarks_packages} shows the supported features of our proposed OMB-Py design compared to three MPI benchmarks packages: 1) mpi4py sample applications~\cite{mpi4py_git}, 2) Intel MPI Benchmarks (IMB)~\cite{imb_bench}, and 3) Sandia MPI Micro-Benchmark Suite (SMB)~\cite{smb}.
\begin{table}
\caption{Feature comparison between the proposed OMB-Py design and other MPI micro-benchmark packages}
\vspace{-2.0ex}
\centering
\begin{tabular}{|l|c|c|c|c|} 
\hline
\rowcolor[rgb]{1,0.855,0.592}                                         & \begin{tabular}[c]{@{}>{\cellcolor[rgb]{1,0.855,0.592}}c@{}}OMB-Py \\(Proposed\\Design)\end{tabular} & \begin{tabular}[c]{@{}>{\cellcolor[rgb]{1,0.855,0.592}}c@{}}mpi4py Sample\\Applications~\cite{mpi4py}\end{tabular} & \begin{tabular}[c]{@{}>{\cellcolor[rgb]{1,0.855,0.592}}c@{}}IMB\\~\cite{imb}\end{tabular} & \begin{tabular}[c]{@{}>{\cellcolor[rgb]{1,0.855,0.592}}c@{}}SMB\\~\cite{smb}\end{tabular}  \\ 
\hline
Point-to-Point                                                        &\ding{51}                                                                     &\ding{51}             &\ding{51}     &\ding{51}      \\ 
\hline
Blocking Collectives                                                  &\ding{51}                                                                     &Partially             &\ding{51}     &\ding{55}     \\ 
\hline
\begin{tabular}[c]{@{}l@{}}Vector Variant \\Blocking Collectives\end{tabular} &\ding{51}                                                                     &Partially             &\ding{51}     &\ding{55}      \\ 
\hline
Support for Python                                                    &\ding{51}                                                                     &\ding{51}             &\ding{55}     &\ding{55}      \\ 
\hline
Bytearray Buffers                                                    &\ding{51}                                                                     &\ding{55}             &\ding{55}     &\ding{55}      \\ 
\hline
Numpy~Buffers                                                         &\ding{51}                                                                     &\ding{51}             &\ding{55}     &\ding{55}      \\ 
\hline
CuPy~Buffers                                                          &\ding{51}                                                                     &\ding{55}             &\ding{55}     &\ding{55}      \\ 
\hline
PyCUDA~Buffers                                                        &\ding{51}                                                                     &\ding{55}             &\ding{55}     &\ding{55}      \\
\hline
\end{tabular}
\label{table:benchmarks_packages}
\vspace{-4.0ex}
\end{table} We use OMB-Py to conduct performance evaluation to gain a better understanding of MPI performance with Python on several HPC clusters. We use OMB to evaluate MPI performance in C to provide a baseline for comparison. This paper makes the following key contributions:
\begin{itemize}
  \item Design and implementation of OMB-Py, a comprehensive micro-benchmarks package to evaluate MPI performance in Python. To the best of our knowledge, OMB-Py is the first communication micro-benchmark for Python MPI applications.
  \item Evaluation on four HPC systems using a variety of point-to-point and collective communication tests for CPU and GPU data-structures/libraries including NumPy, ByteArrays, CuPy, Numba, and PyCUDA.
  \item Evaluation shows a small overhead in latency for MPI operations in Python compared to C. For CPU-based benchmarks, we observed 30\% and 3\% average overhead in latency.
  \item Evaluation shows that CuPy and PyCUDA as GPU-aware data buffers give the best MPI communication performance compared to Numba which shows more overhead. The latency overhead for communicating Numba-based data was almost 2x compared to CuPy and PyCUDA.
  \item Analysis shows that 80\%-90\% of the overhead of mpi4py over native MPI libraries comes from preparing the send and receive buffers to link the Python objects in the Cython layer.
  \item Planned release of the OMB-Py package with benchmarks support for point-to-point and blocking collectives MPI operations.
%  \item To the best of our knowledge, this is the first comprehensive MPI micro-benchmarks package that supports Python.
\end{itemize}

%\subsection{Insight Gained by Evaluating the Proposed Design}
%In the evaluation section of this paper, we conduct extensive experiments on 3 HPC clusters using the proposed OMB-Py design to evaluate performance for point-to-point and collective benchmarks on both CPUs and GPUs. We also evaluate the performance for the pickle method in mpi4py and the distributed ML implementations. 

%Some results insight provided by Nawras: 
%
%    CPU inter-node Frontera numbers
%	
%        Latency: small message sizes: 30\% overhead on average, large message sizes: 3\% overhead on average	
%        BW: small 65\%, large 6\%
%
%    GPU on RI2 (we only reported latency):
%        small: 134\% for cupy, 130\% for pycuda, 225\% for numba
%        large: 13\% for cupy, 13\% for pycuda, 18\% for numba
% \input{text/motivation}
% \input{text/motivation}
\section{Background}
\label{sec:bgnd}

\subsection{OSU Micro Benchmarks}
\label{sec:bgnd:omb}
OSU micro benchmarks (OMB)~\cite{OMB, OMB-GPU} is a widely used package to measure the performance of MPI implementations on HPC systems with different configurations and hardware. It offers a variety of benchmarks to report different MPI communication metrics including point-to-point, blocking/non-blocking collectives, one-sided MPI operations. It provides users with a wide range of options to run customizable tests like the number of iterations per test, message sizes, communication devices, and more. OMB supports a variety of different interfaces like ROCm and CUDA to run on ARM and NVIDIA GPUs. This package is written in C so it can call MPI operations directly.

\subsection{MPI for Python}
The MPI standard mainly defines official language bindings for C and Fortran. A middleware (or wrapper) is needed as a bridge between a high-level programming language like Python and MPI implementations. There are a number of implementations that enable Python to utilize MPI functionalities like mpi4py~\cite{mpi4py}, torch.distributed~\cite{torch.distributed}, pypar, and pyMPI. mpi4py is one of the most widely used wrappers as it offers continuous support and compatibility for newer MPI and Python updates. It also supports a number of Python objects as memory buffers for communication like built-in Python arrays and bytearrays or third-party data structures like NumPy arrays and mmap (memory-mapped file objects). More recently, mpi4py added support for GPU-aware data structures like CuPy, PyCUDA, and Numba. If a data structure is not supported, the mpi4py library offers a variation of its functions to serialize/unserialize the communicated Python object. This is mainly referred to as “pickling” when an object is converted into a byte stream and “unpickling” when it is converted back to its original format. In mpi4py, the MPI methods that use pickle are defined with a lower case first letter such as \underline{s}end(), \underline{r}ecv(), \underline{r}educe(), \underline{a}llgather, etc. The direct buffer methods are defined with upper case first letter such as \underline{S}end(), \underline{R}ecv(), \underline{R}educe(), \underline{A}llgather(), etc.
\section{Proposed Design and Implementation}
\label{sec:proposed_desing}

\begin{figure}[!htbp]
    \centering
    \includegraphics[width=0.75\columnwidth]{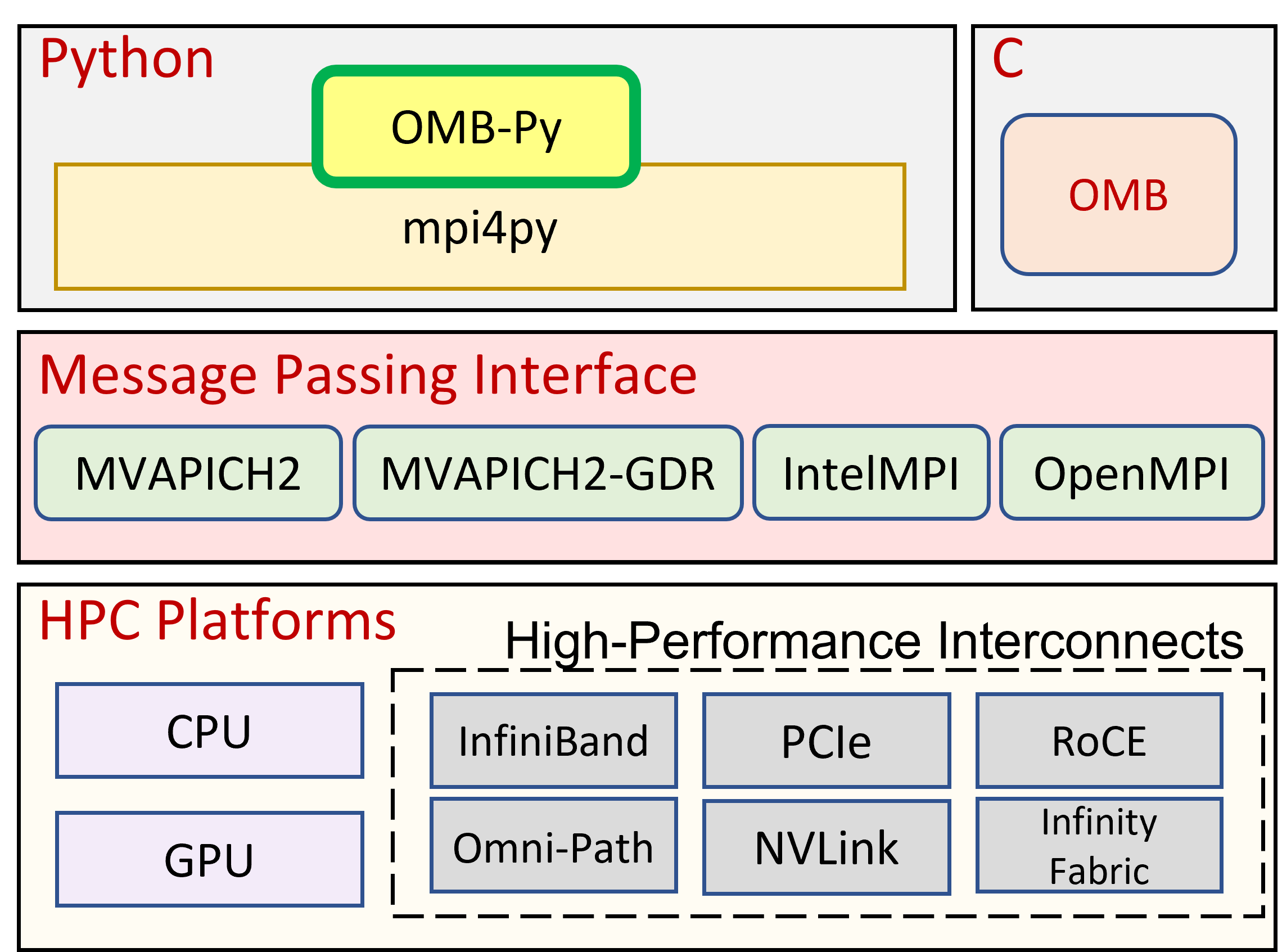}
    \Mycaption{Architectural hierarchy of OMB-Py with mpi4py, MPI, and HPC platforms.}
    \label{fig:layers}
    \vspace{-2.0ex}
\end{figure}

The proposed design OMB-Py is aimed to evaluate performance for MPI communication in the Python programming language using Python objects for communication.  We use the mpi4py package which is widely used to provide Python bindings for the MPI standard. Figure~\ref{fig:layers} shows the architectural hierarchy of OMB-Py with MPI and HPC Platforms. As shown in the figure, OMB-Py needs mpi4py to interact with the MPI layer, whereas OMB can directly interact with MPI as it is written in C.

\subsection{Supported Benchmarks in OMB-Py}
\label{sec:proposed_desing:supported_benchs}
OMB-Py supports benchmarking for a series of MPI operations with a wide range of user options to run customizable tests. Table~\ref{table:benchmarks} shows the point-to-point, blocking collectives, and vector variants benchmarks to be supported in the first release of OMB-Py. The package supports CPU benchmarking using Python built-in bytearrays and NumPy~\cite{numpy} arrays as data buffers, and GPU benchmarking by using GPU-aware arrays including CuPy~\cite{cupy}, PyCUDA~\cite{pycuda}, and Numba~\cite{numba} as data buffers. The package also supports mpi4py pickle method to serialize communicated objects.

\begin{table}
\caption{Point-to-point, blocking collectives, and vector variant blocking collectives Benchmarks supported by OMB-Py}
\vspace{-2.0ex}
\centering
\begin{tabular}{|l|l|} 
\hline
\rowcolor[rgb]{1,0.82,0.49}  & Supported Benchmarks                                                                                                                  \\ 
\hline
Point-to-Point               & \begin{tabular}[c]{@{}l@{}}Bi-directional bandwidth, bandwidth,\\latency, multi latency\end{tabular}                                  \\ 
\hline
Blocking Collectives         & \begin{tabular}[c]{@{}l@{}}Allgather, Allreduce, Alltoall, \\barrier, bcast, gather, \\reduce\_scatter, reduce, scatter\end{tabular}  \\ 
\hline
\begin{tabular}[c]{@{}l@{}}Vector Variant \\Blocking Collectives\end{tabular} & Allgatherv, Alltoallv, Gatherv, Scatterv                                                                                              \\
\hline
\end{tabular}
\label{table:benchmarks}
\vspace{-4.0ex}
\end{table}

\subsection{Proposed Pipeline for Benchmarking MPI Operations}
\label{sec:proposed_desing:bench_pipeline}

In OMB-Py, we try to maintain the original OMB designs as much as possible while writing the package in Python. The purpose of a benchmark is to simulate the use of certain operations in real applications and report the performance accurately. The two main performance metrics in MPI are latency (measured in microseconds) and bandwidth (measured in GB/s). In OMB-Py, we isolate the MPI operation of interest to avoid unnecessary Python commands or control structure because these may considerably influence the performance results. We also run the measured MPI operations for multiple iterations and find the average, max, and min performance across all participating processes.

\subsection{Example: OMB-Py Latency Benchmark}
\label{sec:proposed_desing:example_bench}

Algorithm~\ref{alg:latency} shows a simple example in pseudocode which measures the latency for blocking send/receive MPI operation in a ping-pong fashion. The sender sends a message and wait for a reply, and the receive receives the message and sends back a reply with the same size. 

\LinesNumbered

\begin{algorithm}
 \label{alg:latency}
 init\_MPI\_communication(...)\;
 allocate(s\_buf, ...)\;
 allocate(r\_buf, ...)\;

 \For{size in message\_sizes}{
  MPI\_Barrier()\;
  \eIf{myrank == 0}{
   start\_time = current\_time()\;
   \For {i: 1 ... max\_iterations}{
   MPI\_Send(s\_buf, size ...)\;
   MPI\_Recv(r\_buf, size ...)\;
   }
   end\_time = current\_time()\;
   latency = (start\_time - end\_time)\;
   }{
   start\_time = current\_time()\;
   \For {i: 1 ... max\_iterations}{
   MPI\_Recv(r\_buf, size ...)\;
   MPI\_Send(s\_buf, size ...)\;
   }
   end\_time = current\_time()\;
   latency = (start\_time - end\_time)\;
  }
  latency = latency / (2 * max\_iterations)\;
  report\_latency()\;
 }
 \caption{Blocking Send/Recv Latency Benchmark}
\end{algorithm}

The latency is averaged across multiple iterations for more accuracy. The placement of MPI\_Barrier() at line 5 guarantees that both sender and receiver processes will start their operations at the same time. This general pipeline is followed throughout the different benchmarks; however, some specifics will differ from benchmark to another depending on the measured metric, communicated datatypes, number of participating processes, and devices used to carry out the communication. Moreover, with collective benchmarks we need to find the average latency across all participating processes; thus, we use MPI\_Reduce() to find that average then report the latency.

\subsection{CPU Memory Buffer Datatypes}
\label{sec:proposed_desing:bench_cpu}

While maintaining the benchmarking pipeline explained earlier, there is still a number of design options for implementing MPI benchmarks in Python. mpi4py supports both built-in Python objects and third-part libraries as communication buffers. For CPUs, we choose to add support for the following data structures: 1) Built-in bytearrays and 2) NumPy arrays. 

\subsection{GPU Memory Buffer Datatypes}

mpi4py recently added support for Python GPU-aware libraries as data buffers~\cite{mpi4py_12y}. Those libraries have defined a protocol called CUDA Aware Interface (CAI)~\cite{cai} which requires CUDA array-like objects to add a new Python attribute that contains a pointer to the GPU buffer address. This protocol guarantees interoperability between different implementations of the GPU-aware libraries in Python. For OMB-Py, we choose to add support for three of these libraries as buffers for GPU communication: 1) CuPy~\cite{cupy}, 2) Numba~\cite{numba}, and 3) PyCuda~\cite{pycuda}. Similar to NumPy, each of these libraries allow initializing different types of arrays and carry out complex matrix operations. When mpi4py is built against CUDA-aware MPI, those arrays can be passed to MPI operations calls.

\subsection{User Options for OMB-Py}
In this subsection, we explain the different options that OMB-Py users modify to run custom tests.
\begin{itemize}
  \item Device: can choose either CPU or GPU devices to run the experiments on.
  \item Buffer: can choose from a list of Python objects to use as buffers. The list includes: bytearrays, Numpy, CuPy, PyCUDA, and Numba arrays.
  \item Message size: defines lower and upper limits for message sizes to report performance for.
  \item Number of iterations: defines number of times the tested MPI operation is executed. Reported performance numbers are the overall averages of all runs.
  \item Number of warm-up iterations: defines the number of times to run the MPI operation before starting the actual test.
\end{itemize}
\section{Evaluation}
\label{sec:evaluation}

This section provides a comprehensive evaluation of the proposed OMB-Py benchmarks on various HPC clusters. We use MVAPICH2~\cite{mvapich2} for the CPU tests and MVAPICH2-GDR for the GPU-aware tests. We also use OMB benchmarks in C as a point of reference to evaluate the benchmarks of OMB-Py. First, we describe the different experimental environments we used for evaluation. We present the evaluation results for point-to-point and collective tests on CPU and then on GPU using three GPU-aware Python memory buffers 1) CuPy~\cite{cupy}, 2) PyCUDA~\cite{pycuda}, and 3) Numba~\cite{numba} and compare them to the OMB benchmarks performance. We evaluate the performance of the pickle methods compared to direct buffers in mpi4py.

\subsection{Experimental Setup}
\label{sec:evaluation:setup}

The CPU experiments are performed on four different clusters. Here are the HPC clusters we used: 
\subsubsection{Frontera}
Frontera is the largest NSF funded HPC system. It is deployed and maintained at the Texas Advanced Computing Center (TACC). On this cluster, we perform experiments on up to 16 Intel x86 compute nodes that have the Intel Xeon Platinum 8280 (Cascade Lake) processors. Each node has two sockets with 28 cores per socket (56 per node) at 2.70GHz frequency and 192GB of RAM per node. The system is interconnected by Mellanox InfiniBand HDR and HDR-100 interconnect.
\subsubsection{Stampede2}
Stampede2 is also at the Texas Advanced Computing Center. On this cluster, we perform experiments on up to 16 Skylake nodes with the Intel(R) Xeon(R) Platinum 8160 processors. Each node has two sockets with 24 cores per socket (48 cores per node) and 2 physical threads per core at 2.70GHz frequency and 192GB of RAM per node. The system is interconnected by Intel Omni-Path.
\subsubsection{RI2}
RI2 is an in-house cluster at The Ohio State University. On this cluster, we perform experiments on up to 8 nodes equipped with Intel(R) Xeon(R) Gold 6132 processors with two sockets. Each socket has 14 cores each (28 core per node) at 2.40GHz frequency. This system is interconnected using Mellanox SB7790 and SB7800 InfiniBand switches.

\subsubsection{Bridges-2 (GPU)}
Bridges-2 is a supercomputer system at the Pittsburgh Supercomputing Center. All GPU experiments are performed on the Bridges-2 cluster. We use 16 GPUs on 2 nodes. Nodes have Intel Xeon Gold 6248 “Cascade Lake” with 40 cores (20 per socket) at 2.50GHz frequency with 512GB of RAM. Each node has eight NVIDIA Tesla V100-32GB SXM2 GPUs. The system is interconnected by Mellanox Infiniband and every node has two Mellanox ConnectX-6 HDR Infiniband 200Gb/s Adapters. 

\subsection{Used Software Packages}
For the CPU experiments on all four HPC systems, we use MVAPICH2 2.3.6~\cite{mvapich2} for MPI and OMB v5.8~\cite{OMB} as a point of reference. We also use mpi4py~\cite{mpi4py} v3.1.1 built against MVAPICH2 as a wrapper for Python bindings with MPI.
For the GPU experiments, we use MVAPICH2-GDR 2.3.6 built against CUDA 11.2 for MPI and OMB v5.8 as a point of reference. We use mpi4py v3.1.1 built against CUDA 11.2 and MVAPICH2-GDR as a wrapper for Python bindings with MPI.

\subsection{Point-to-Point Intra-node Evaluation on CPU}
\label{sec:evaluation:intra_cpu}
This subsection provides intra-node CPU performance evaluation for the three experimental setups on Frontera, Stampede2, and RI2.

\subsubsection{Frontera}

\begin{figure}[!htbp]
    \centering
    \includegraphics[width=0.75\columnwidth]{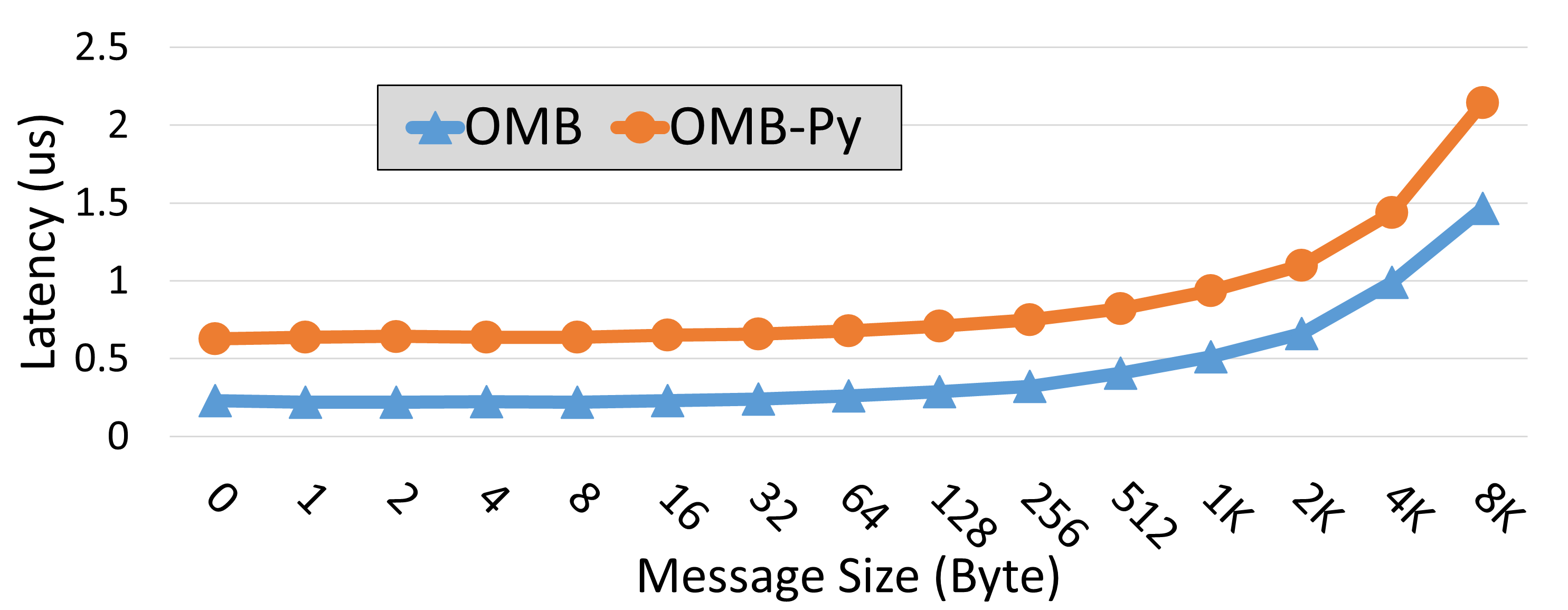}
    \Mycaption{Intra-node CPU latency for small message sizes comparing OMB-Py and OMB benchmarks on the Frontera cluster.}
    \label{fig:latency_frontera_small}
    \vspace{-1ex}
\end{figure}

\begin{figure}[!htbp]
    \centering
    \includegraphics[width=0.75\columnwidth]{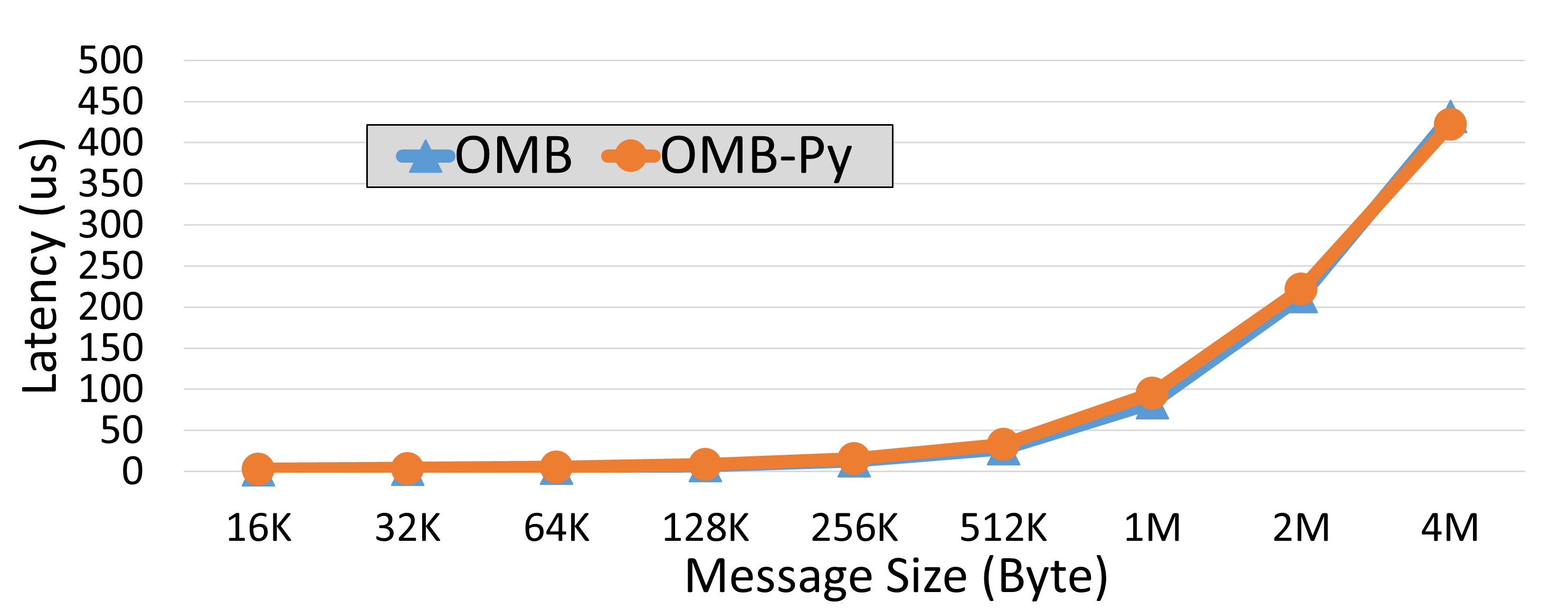}
    \Mycaption{Intra-node CPU latency for large message sizes comparing OMB-Py and OMB benchmarks on the Frontera cluster.}
    \label{fig:latency_frontera_large}
    \vspace{0.5ex}
\end{figure}

Figure~\ref{fig:latency_frontera_small} shows the intra-node latency curves for OMB and OMB-Py for small message sizes on Frontera. Both latency curves follow the same trend; however, OMB-Py latency numbers have an average overhead of 0.44 microseconds compared to OMB numbers.
Figure~\ref{fig:latency_frontera_large} shows latency numbers for the same benchmark but for larger message sizes. Although the two curves for OMB and OMB-Py are almost identical, there's a relatively small overhead of 2.31 microseconds on average for this message size range.

\subsubsection{Stampede2}
\begin{figure}[!htbp]
    \centering
    \includegraphics[width=0.75\columnwidth]{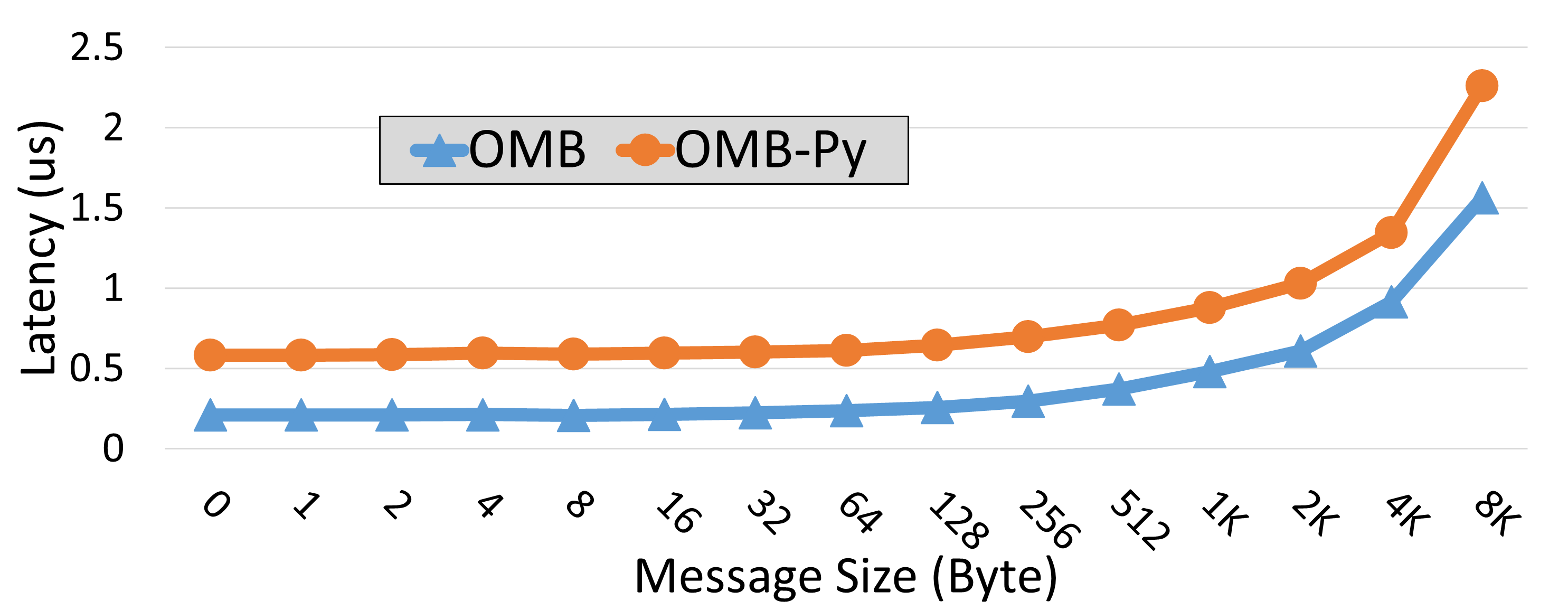}
    \Mycaption{Intra-node CPU latency for small message sizes comparing OMB-Py and OMB benchmarks on the Stampede2 cluster.}
    \label{fig:latency_stampede2_small}
    \vspace{-0.5ex}
\end{figure}

\begin{figure}[!htbp]
    \centering
    \includegraphics[width=0.75\columnwidth]{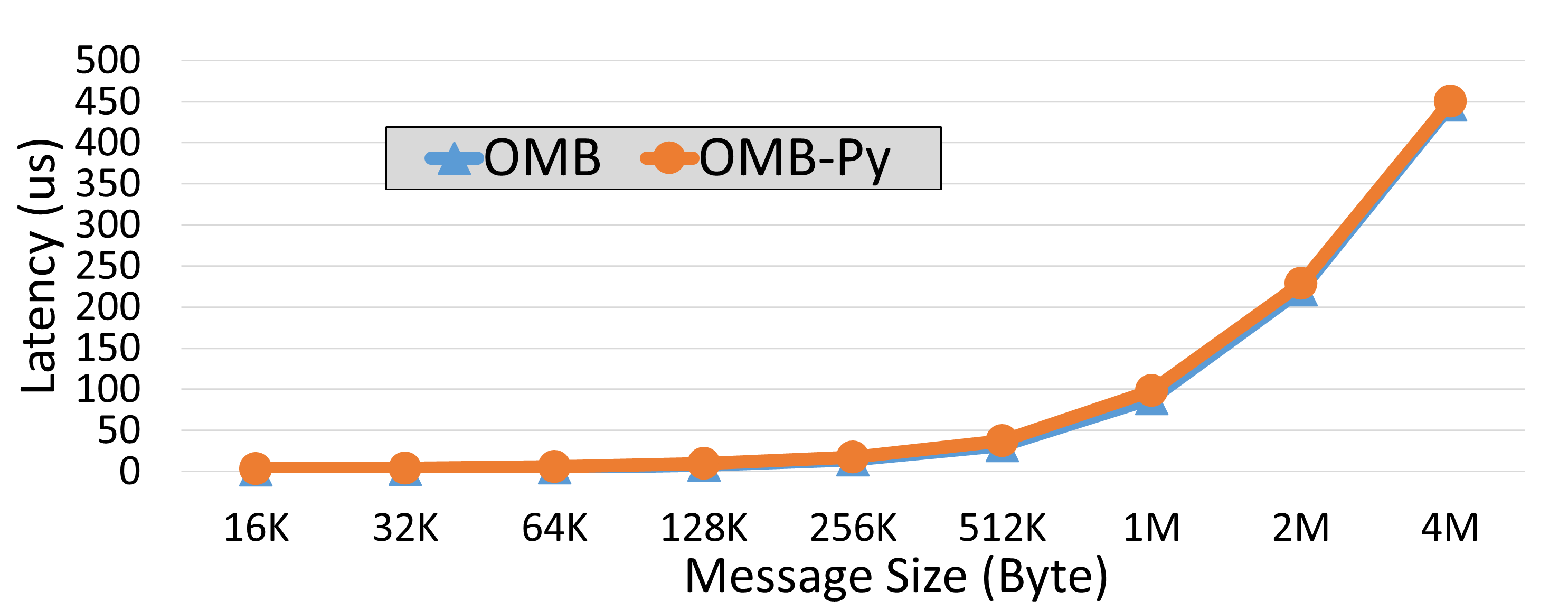}
    \Mycaption{Intra-node CPU latency for large message sizes comparing OMB-Py and OMB benchmarks on the Stampede2 cluster.}
    \label{fig:latency_stampede2_large}
    \vspace{-0.5ex}
\end{figure}

Figure~\ref{fig:latency_stampede2_small} shows the intra-node latency curves for OMB and OMB-Py for small message sizes on Stampede2. Both latency curves follow the same trend; however, OMB-Py latency numbers have an average overhead of 0.41 microseconds compared to OMB numbers.
Figure~\ref{fig:latency_stampede2_large} shows latency numbers for the same benchmark but for larger message sizes. Although the two curves for OMB and OMB-Py are almost identical, there's a relatively small overhead of 4.13 microseconds on average for this message size range.

\subsubsection{RI2}
\begin{figure}[!htbp]
    \centering
    \includegraphics[width=0.75\columnwidth]{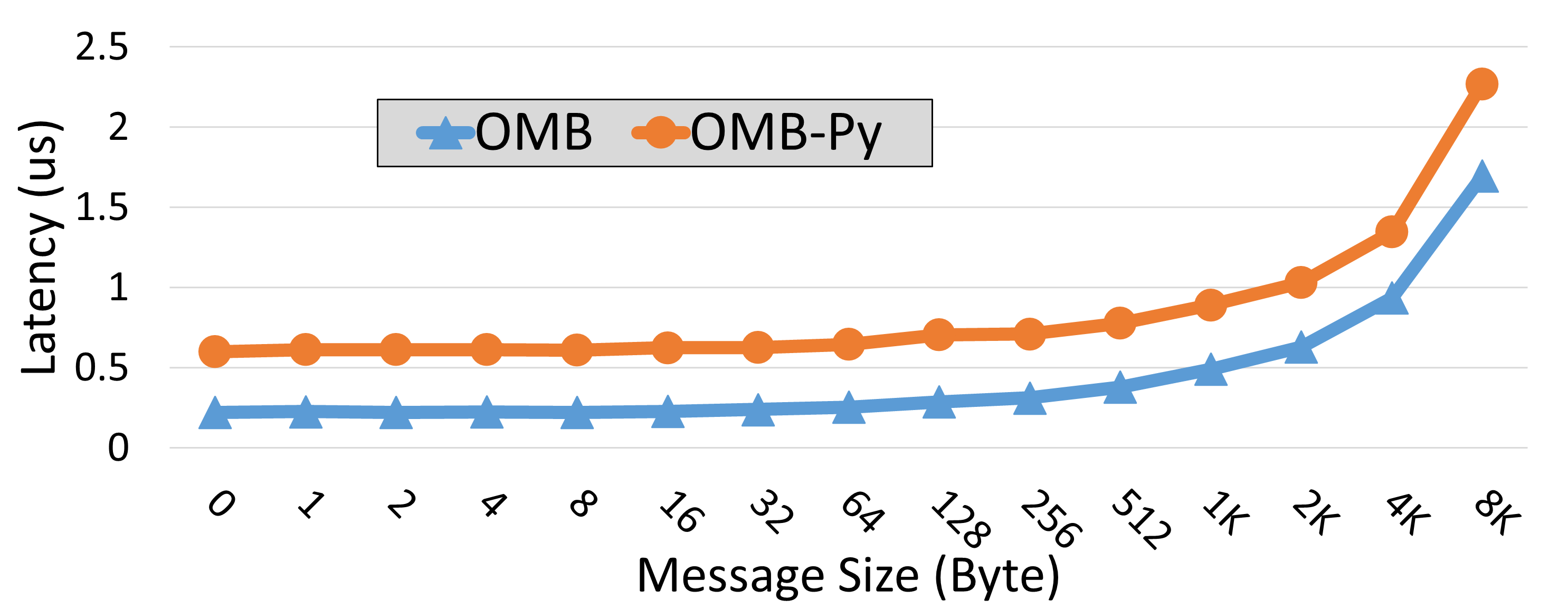}
    \Mycaption{Intra-node CPU latency for small message sizes comparing OMB-Py and OMB benchmarks on the RI2 cluster.}
    \label{fig:latency_ri2_small}
    \vspace{-0.5ex}
\end{figure}

\begin{figure}[!htbp]
    \centering
    \includegraphics[width=0.75\columnwidth]{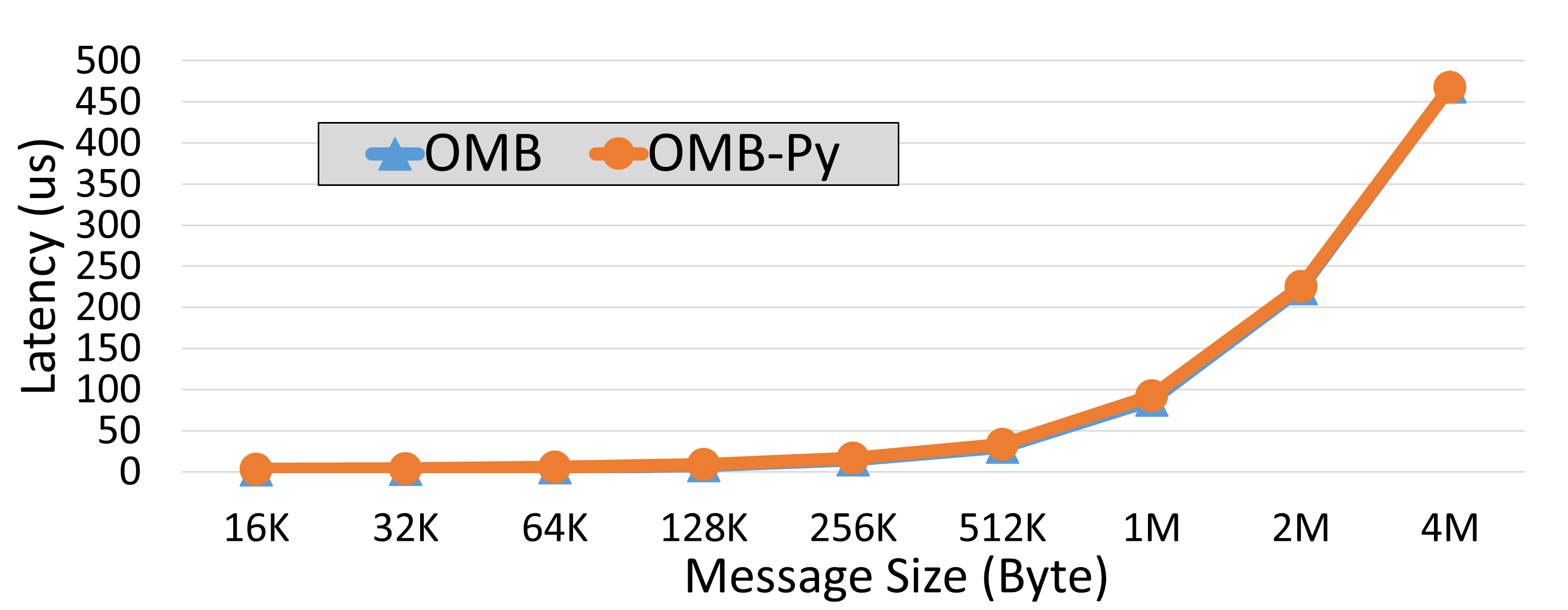}
    \Mycaption{Intra-node CPU latency for large message sizes comparing OMB-Py and OMB benchmarks on the RI2 cluster.}
    \label{fig:latency_ri2_large}
    \vspace{0.5ex}
\end{figure}

Figure~\ref{fig:latency_ri2_small} shows the intra-node latency curves for OMB and OMB-Py for small message sizes on RI2. Both latency curves follow the same trend; however, OMB-Py latency numbers have an average overhead of 0.41 microseconds compared to OMB numbers.
Figure~\ref{fig:latency_ri2_large} shows latency numbers for the same benchmark but for larger message sizes. Although the two curves for OMB and OMB-Py are almost identical, there's a relatively small overhead of 1.76 microseconds on average for this message size range.

Latency numbers on all three platforms follow the same trend. OMB-Py always has an overhead compared to OMB. This overhead is more noticeable in the smaller message size range. For larger message sizes, the overhead is still there but it is relatively smaller. For the rest of the CPU experiments, we perform them mainly on the Frontera system.

\subsection{Point-to-Point Inter-node Evaluation on CPU}
In this section, we conduct evaluation for inter-node point-to-point communication on the Frontera cluster. We present latency and bandwidth numbers for OMB-Py and OMB.

\label{sec:evaluation:inter_cpu}
\subsubsection{Latency}
\begin{figure}[!htbp]
    \centering
    \includegraphics[width=0.75\columnwidth]{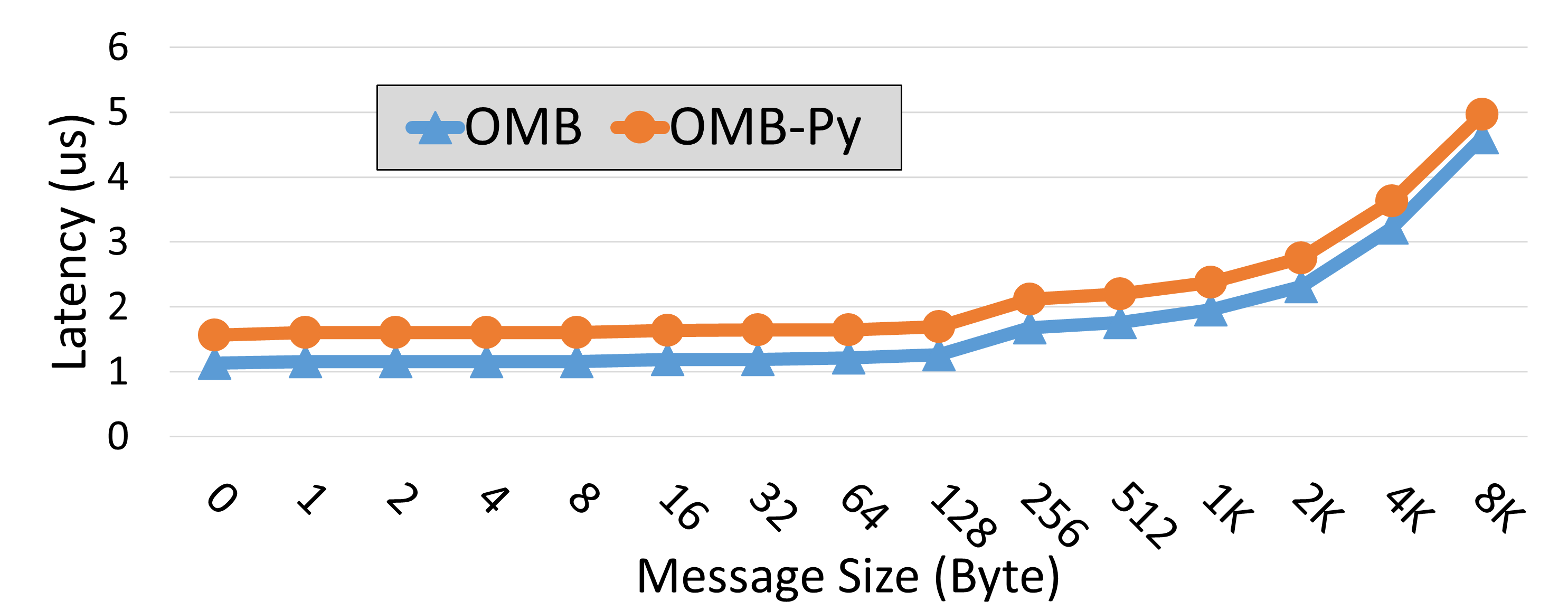}
    \Mycaption{Inter-node CPU latency for small message sizes comparing OMB-Py and OMB benchmarks on the Frontera cluster.}
    \label{fig:inter_latency_frontera_small}
    \vspace{-1ex}
\end{figure}

\begin{figure}[!htbp]
    \centering
    \includegraphics[width=0.75\columnwidth]{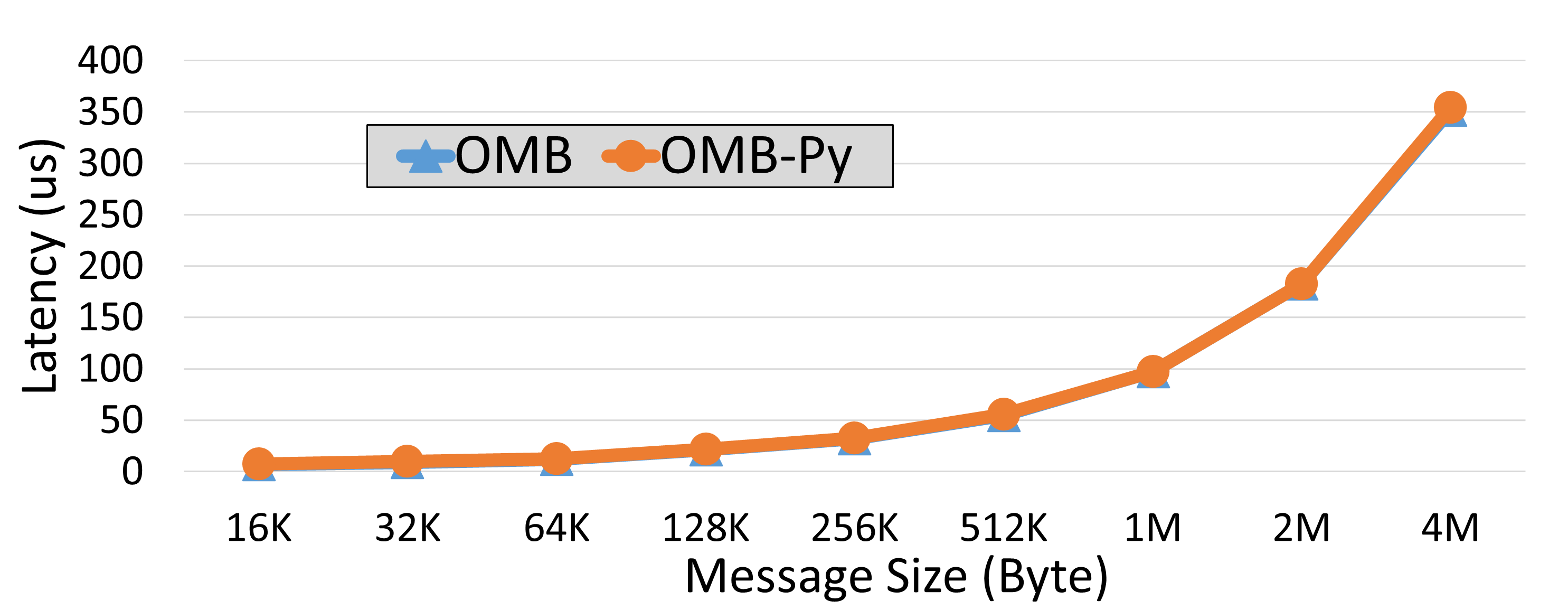}
    \Mycaption{Inter-node CPU latency for large message sizes comparing OMB-Py and OMB benchmarks on the Frontera cluster.}
    \label{fig:inter_latency_frontera_large}
    \vspace{-0.5ex}
\end{figure}

Figure~\ref{fig:inter_latency_frontera_small} shows the inter-node latency curves for OMB and OMB-Py for small message sizes on Frontera. Both latency curves follow the same trend; however, OMB-Py latency numbers have an average overhead of 0.43 microseconds compared to OMB numbers.
Figure~\ref{fig:inter_latency_frontera_large} shows latency numbers for the same benchmark but for larger message sizes. Although the two curves for OMB and OMB-Py are almost identical, there's a small overhead of 0.63 microseconds on average for this message size range.

\subsubsection{Bandwidth}
\begin{figure}[!htbp]
    \centering
    \includegraphics[width=0.75\columnwidth]{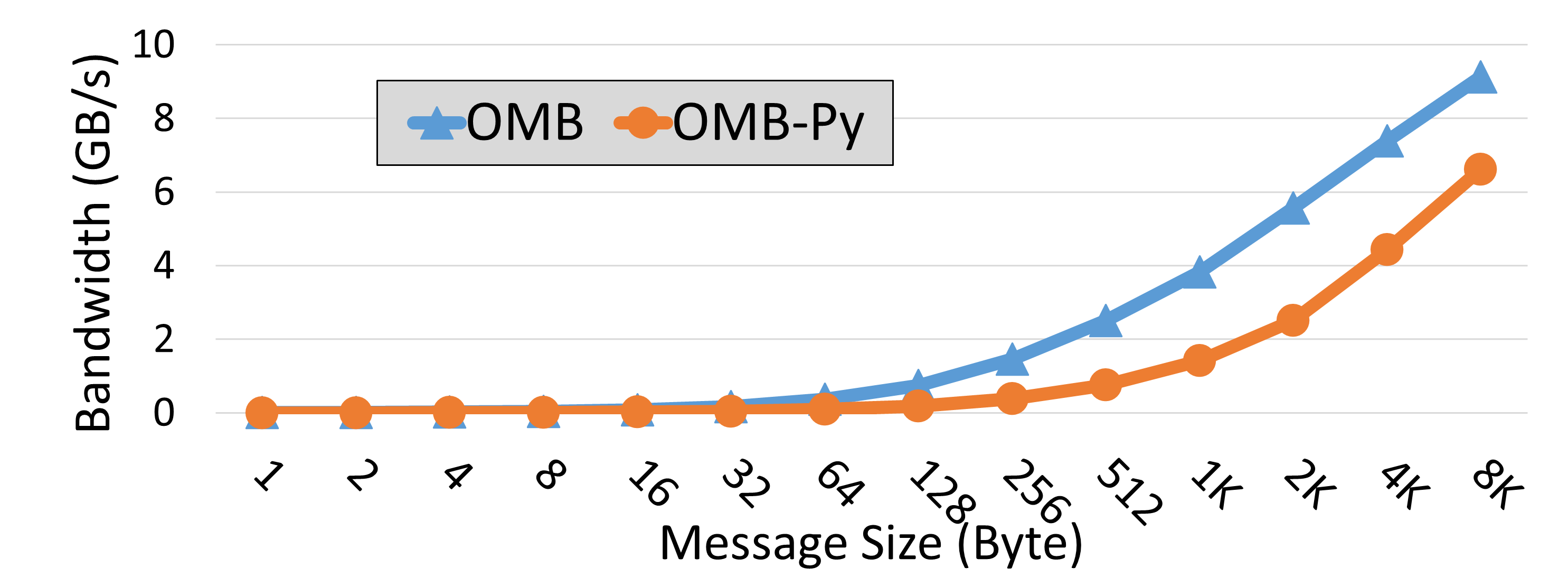}
    \Mycaption{Inter-node CPU bandwidth for small message sizes comparing OMB-Py and OMB benchmarks on the Frontera cluster.}
    \label{fig:inter_bw_frontera_small}
    \vspace{-1ex}
\end{figure}

\begin{figure}[!htbp]
    \centering
    \includegraphics[width=0.75\columnwidth]{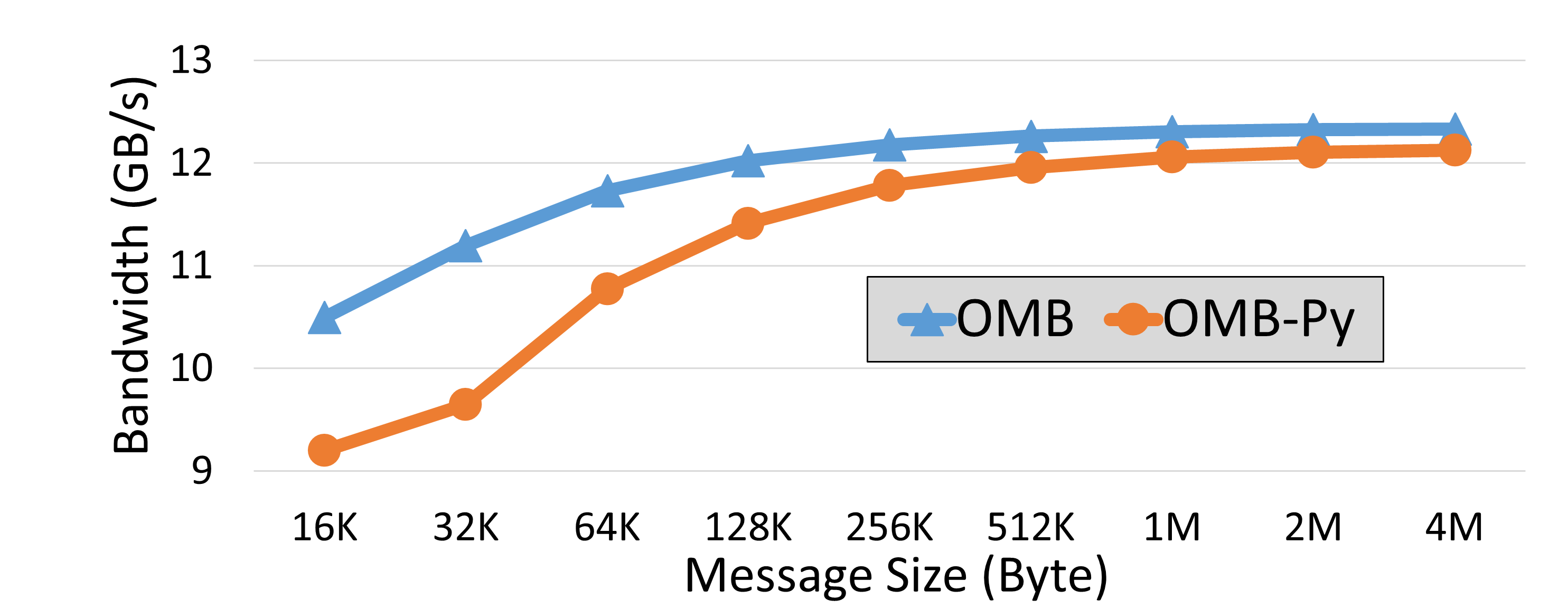}
    \Mycaption{Inter-node CPU bandwidth for large message sizes comparing OMB-Py and OMB benchmarks on the Frontera cluster.}
    \label{fig:inter_bw_frontera_large}
    \vspace{-0.5ex}
\end{figure}

Figure~\ref{fig:inter_bw_frontera_small} shows the inter-node bandwidth curves in GB/s for OMB and OMB-Py for small message sizes on Frontera. Bandwidth for small message sizes (up to 32B) looks similar; however, OMB-Py numbers start to have an average overhead of 1.05GB/s for message sizes 512B to 8KB.
Figure~\ref{fig:inter_bw_frontera_large} shows bandwidth numbers for the same benchmark but for larger message sizes. For this message size range, the OMB-Py overhead starts to shrink again to reach an average overhead of 331MB/s only.

\subsection{Collective Communication Evaluation on CPU}
\label{sec:evaluation:col_cpu}
In this section, we expand on the Frontera results and present numbers for the Allreduce and Allgather operations on 16 nodes for both 1 process per node and 56 processes per node (full subscription).

\subsubsection{Allreduce 1 Process Per Node}
\begin{figure}[!htbp]
    \centering
    \includegraphics[width=0.75\columnwidth]{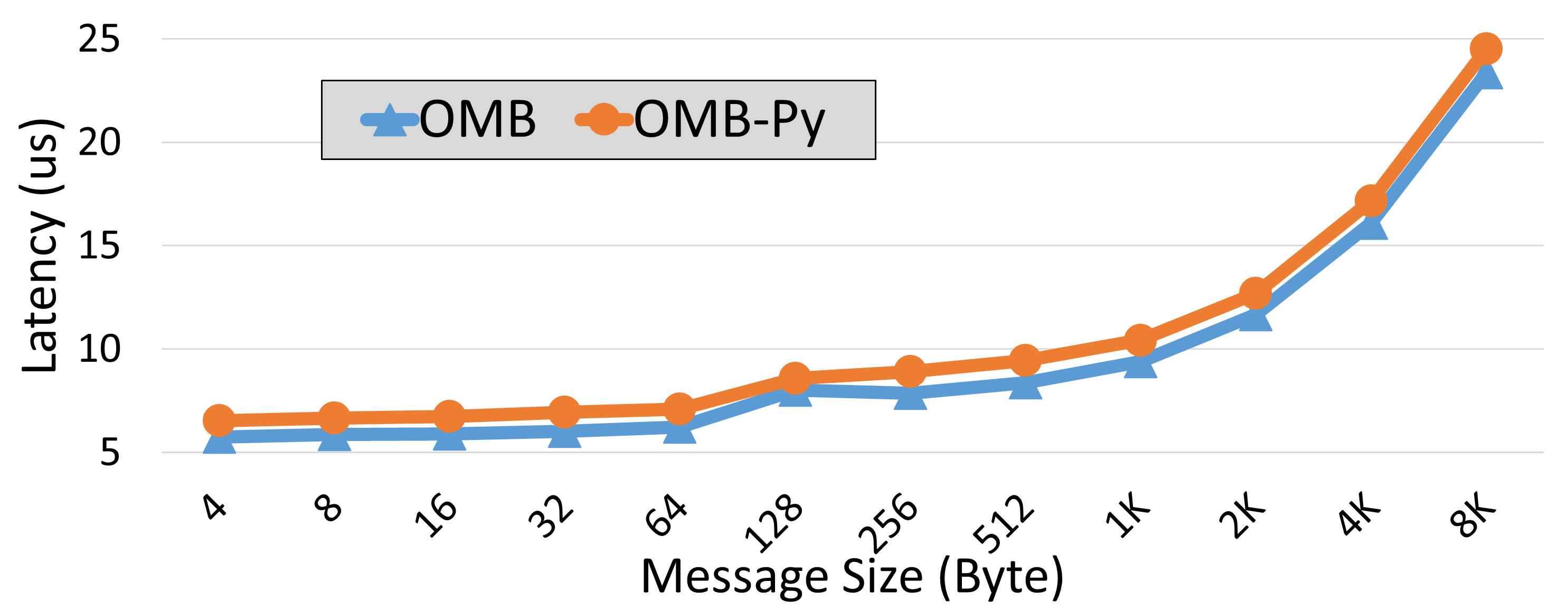}
    \Mycaption{Allreduce CPU latency for small message sizes comparing OMB-Py and OMB benchmarks using 16 nodes and 1 process per node on the Frontera cluster.}
    \label{fig:allreduce_small_1PPN}
    \vspace{-1ex}
\end{figure}

\begin{figure}[!htbp]
    \centering
    \includegraphics[width=0.75\columnwidth]{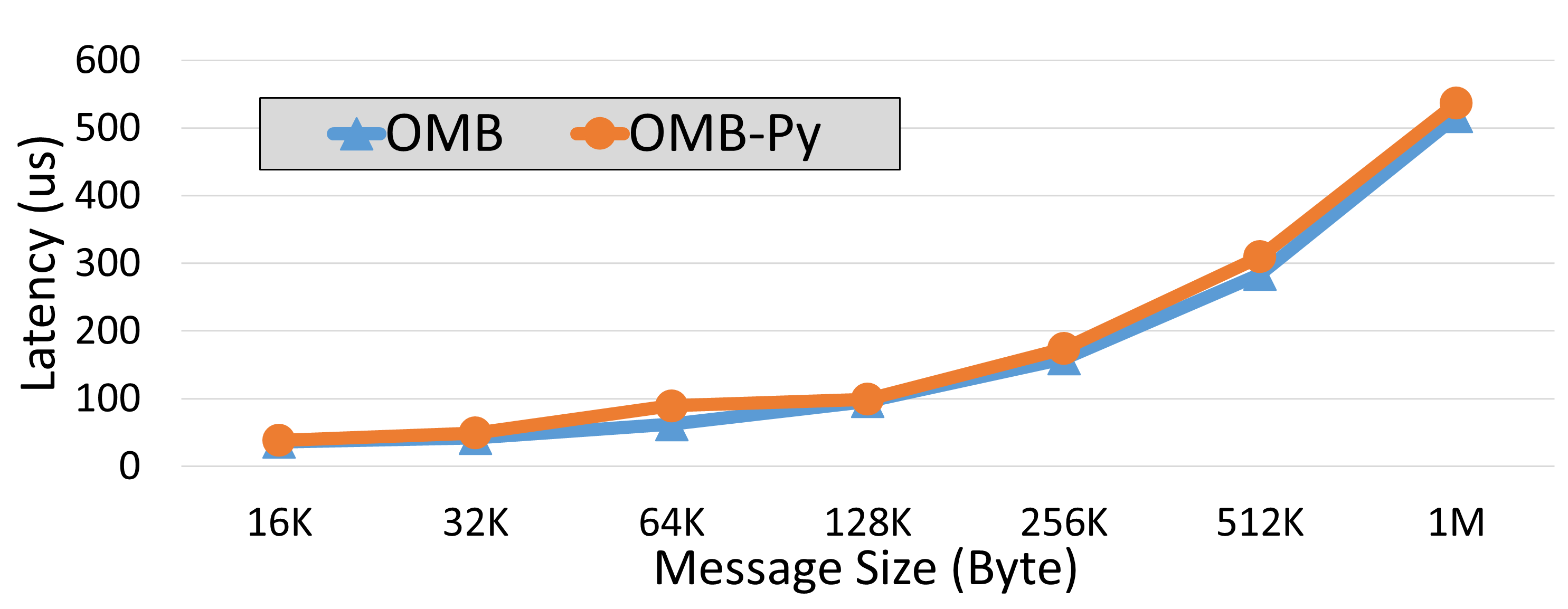}
    \Mycaption{Allreduce CPU latency for large message sizes comparing OMB-Py and OMB benchmarks using 16 nodes and 1 process per node on the Frontera cluster.}
    \label{fig:allreduce_large_1PN}
    \vspace{-0.5ex}
\end{figure}

Figure~\ref{fig:allreduce_small_1PPN} shows the Allreduce latency curves for OMB and OMB-Py for small message sizes with 1 process per node on 16 nodes. OMB-Py has an overhead of 0.93 microseconds for the small message size range.
Figure~\ref{fig:allreduce_large_1PN} shows numbers for the same benchmark but for larger message sizes. The overhead for OMB-Py for the large message size range is 14.13 microseconds on average.

\subsubsection{Allreduce 56 Processes Per Node (Full Subscription)}
\begin{figure}[!htbp]
    \centering
    \includegraphics[width=0.75\columnwidth]{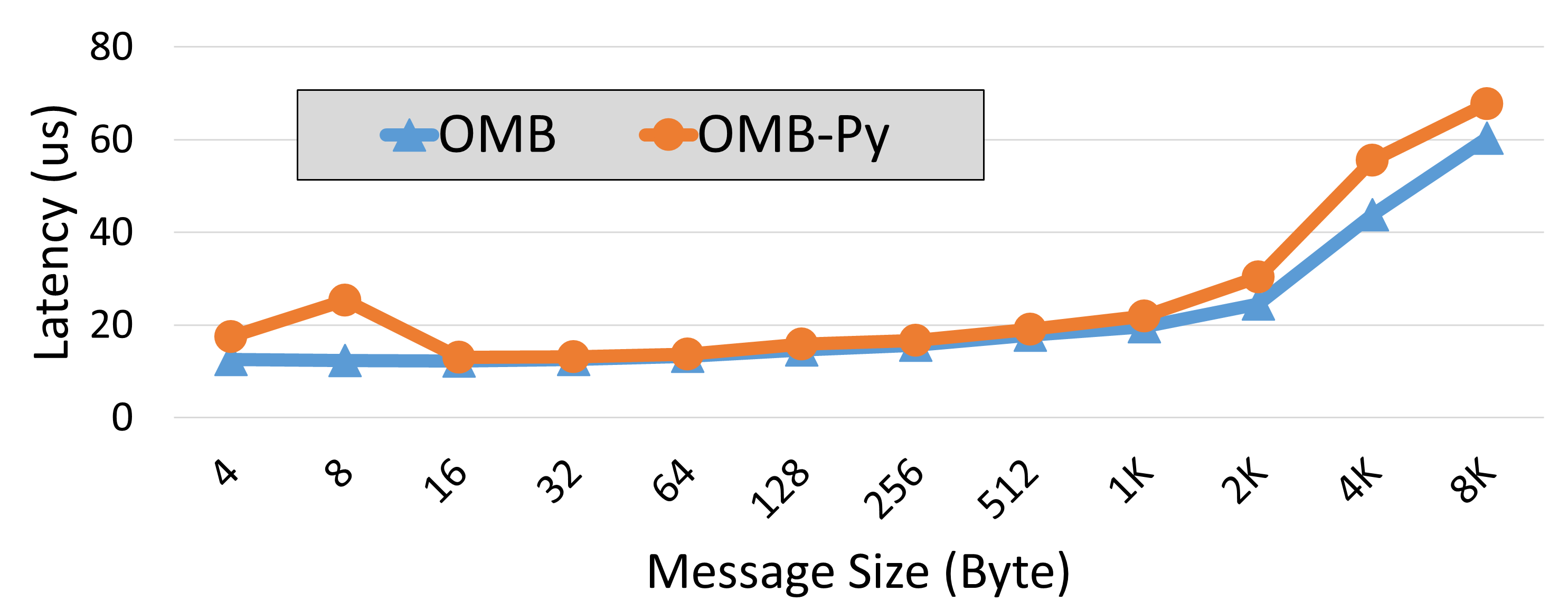}
    \Mycaption{Allreduce CPU latency for small message sizes comparing OMB-Py and OMB benchmarks using 16 nodes and 56 processes per node on the Frontera cluster.}
    \label{fig:allreduce_small_56PPN}
    \vspace{-1ex}
\end{figure}

\begin{figure}[!htbp]
    \centering
    \includegraphics[width=0.75\columnwidth]{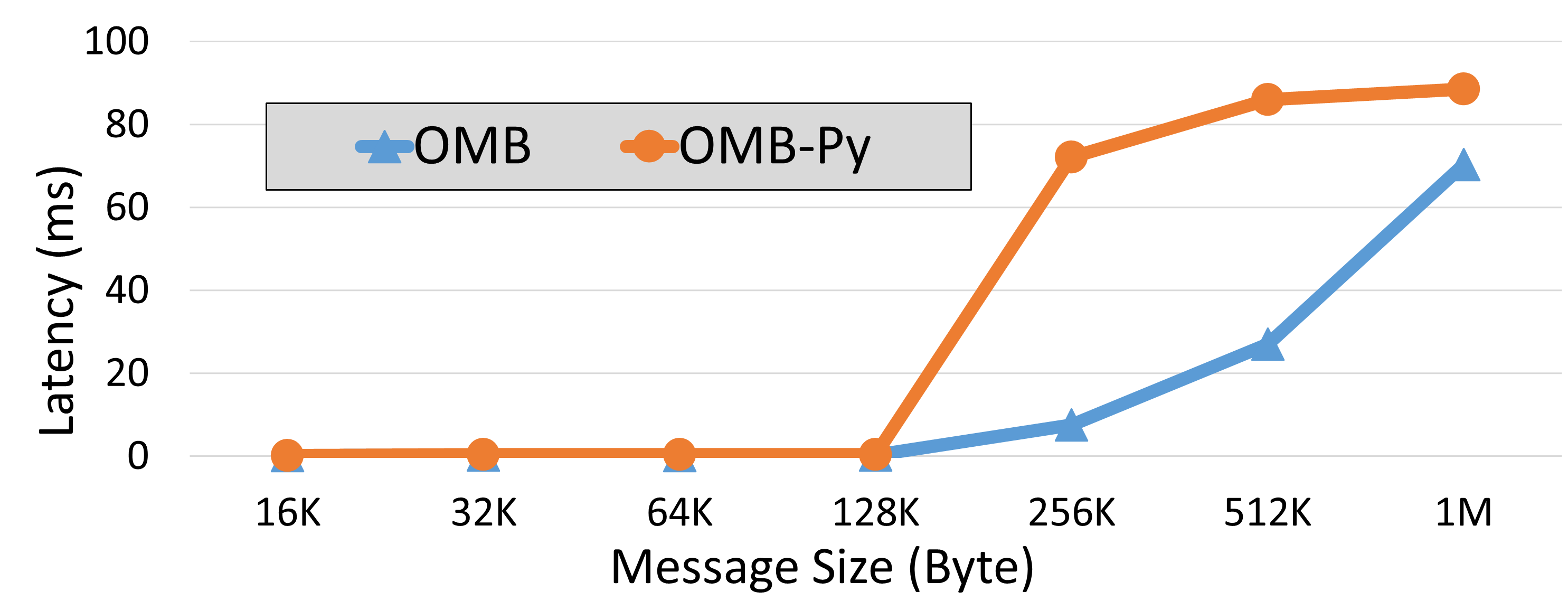}
    \Mycaption{Allreduce CPU latency for large message sizes comparing OMB-Py and OMB benchmarks using 16 nodes and 56 processes per node on the Frontera cluster.}
    \label{fig:allreduce_large_56PPN}
    \vspace{0.5ex}
\end{figure}

Figure~\ref{fig:allreduce_small_56PPN} shows the Allreduce latency curves for OMB and OMB-Py for small message sizes with 56 processes per node (full subscription) on 16 nodes. OMB-Py has an overhead of 4.21 microseconds for the small message size range.
Figure~\ref{fig:allreduce_large_56PPN} shows numbers for the same benchmark but for larger message sizes. OMB initializes MPI with \textit{THREAD\_SINGLE} support in \textit{osu\_latency}; however, default MPI initialization in mpi4py is \textit{THREAD\_MULTIPLE}. This leads to over-subscription of cores in 56 processes per node (full subscription) OMB-Py experiment. Since Allreduce also has computation, over-subscription of threads leads to degradation in large messages for OMB-Py. 

\subsubsection{Allgather 1 Process Per Node}
\begin{figure}[!htbp]
    \centering
    \includegraphics[width=0.75\columnwidth]{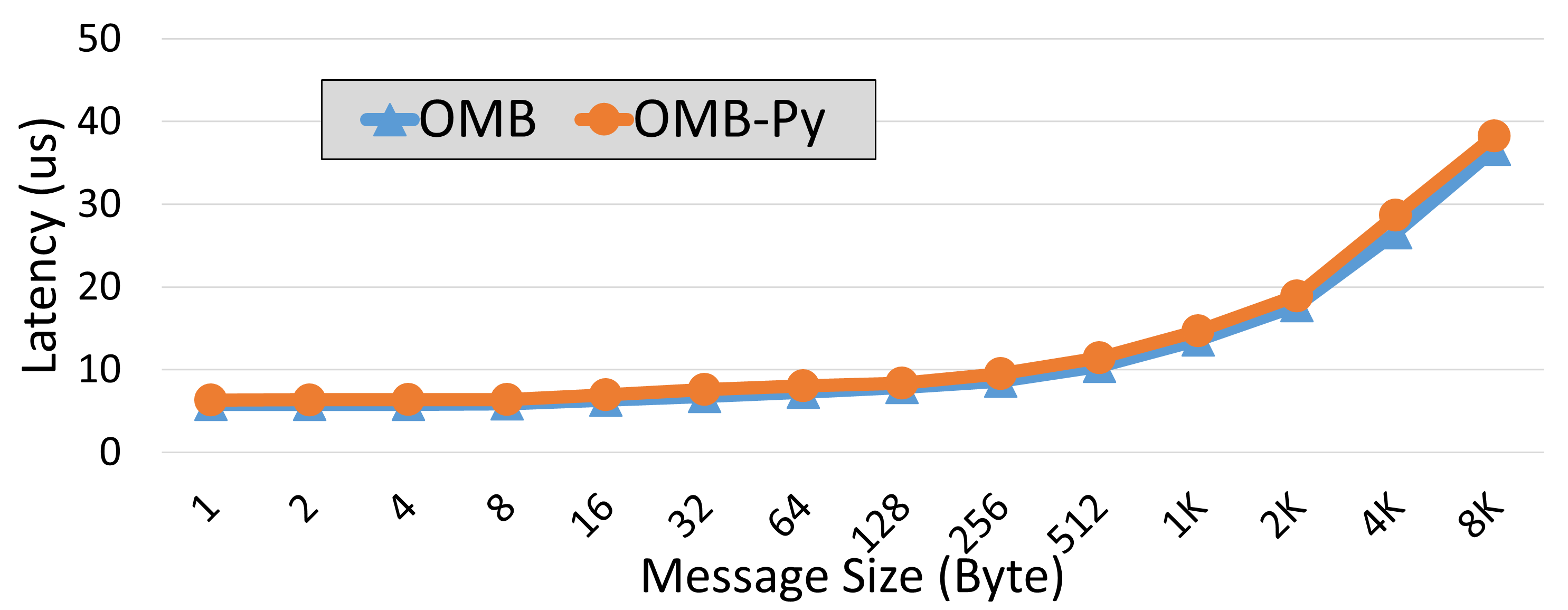}
    \Mycaption{Allgather CPU latency for small message sizes comparing OMB-Py and OMB benchmarks using 16 nodes and 1 process per node on the Frontera cluster.}
    \label{fig:allgather_small_1PPN}
    \vspace{-1ex}
\end{figure}

\begin{figure}[!htbp]
    \centering
    \includegraphics[width=0.75\columnwidth]{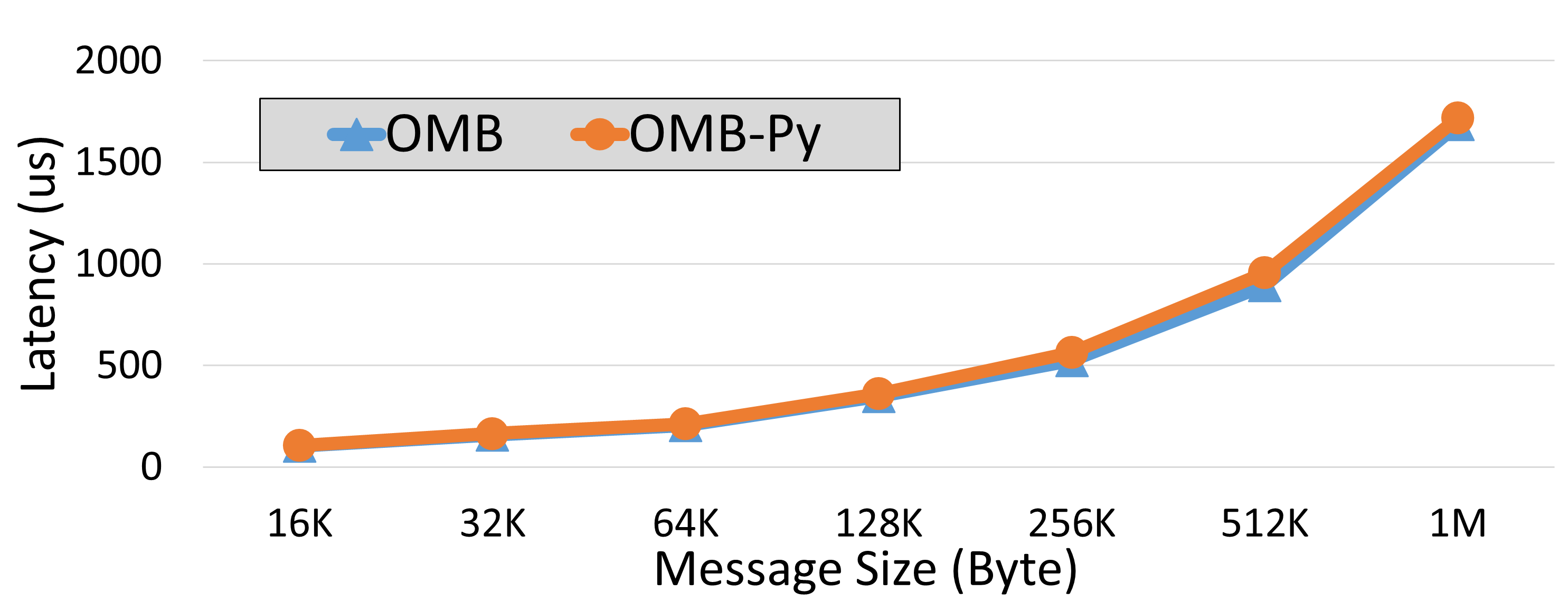}
    \Mycaption{Allgather CPU latency for large message sizes comparing OMB-Py and OMB benchmarks using 16 nodes and 1 process per node on the Frontera cluster.}
    \label{fig:allgather_large_1PN}
    \vspace{0.5ex}
\end{figure}

Figure~\ref{fig:allgather_small_1PPN} shows the Allgather latency curves for OMB and OMB-Py for small message sizes with 1 process per node on 16 nodes. OMB-Py has an overhead of 0.92 microseconds for the small message size range.
Figure~\ref{fig:allgather_large_1PN} shows numbers for the same benchmark but for larger message sizes. The overhead for OMB-Py for the large message size range is 23.4 microseconds on average.

\subsubsection{Allgather 56 Processes Per Node (Full Subscription)}
\begin{figure}[!htbp]
    \centering
    \includegraphics[width=0.75\columnwidth]{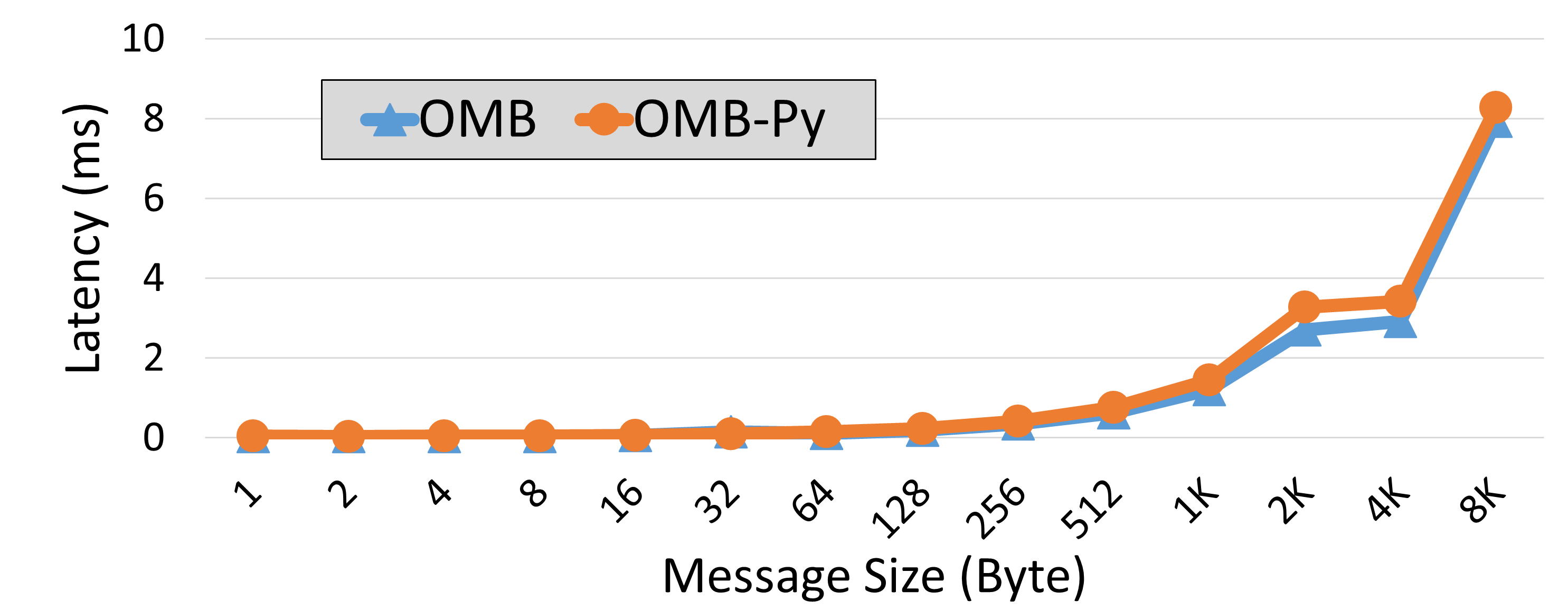}
    \Mycaption{Allgather CPU latency for small message sizes comparing OMB-Py and OMB benchmarks using 16 nodes and 56 processes per node on the Frontera cluster.}
    \label{fig:allgather_small_56PPN}
    \vspace{-1ex}
\end{figure}

\begin{figure}[!htbp]
    \centering
    \includegraphics[width=0.75\columnwidth]{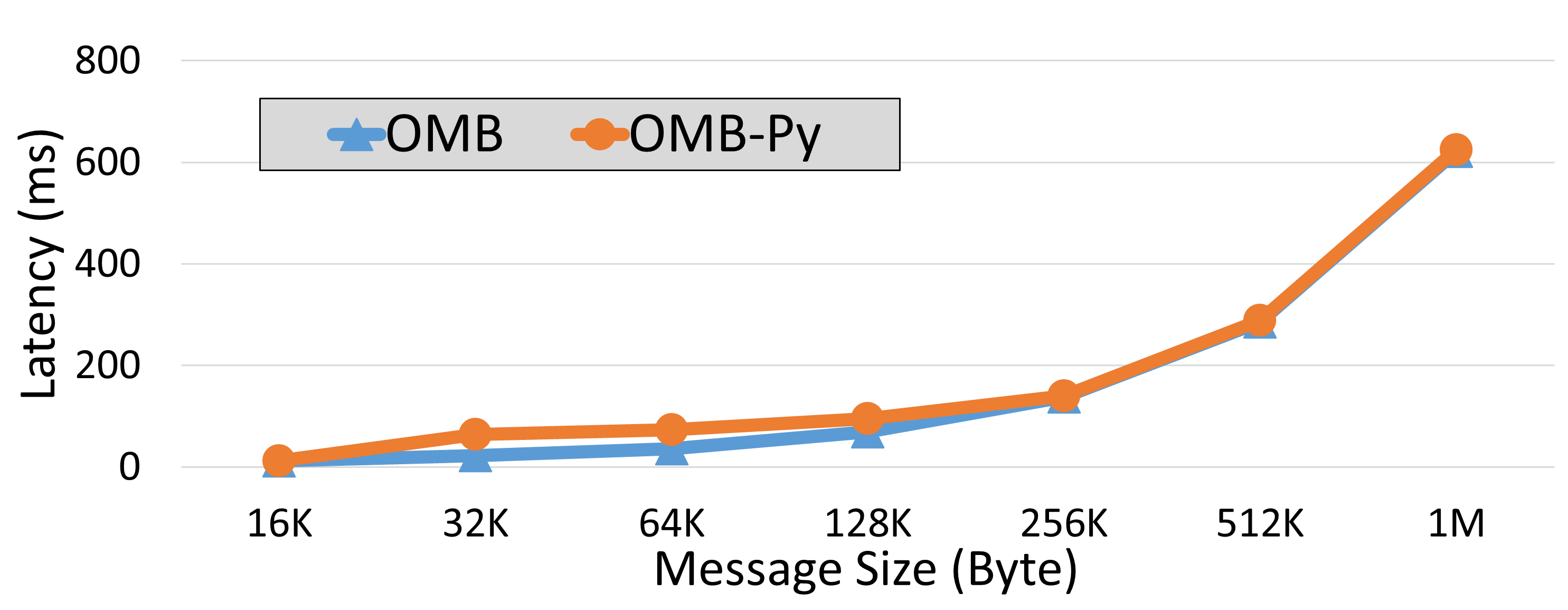}
    \Mycaption{Allgather CPU latency for large message sizes comparing OMB-Py and OMB benchmarks using 16 nodes and 56 processes per node on the Frontera cluster.}
    \label{fig:allgather_large_56PPN}
    \vspace{-0.5ex}
\end{figure}

Figure~\ref{fig:allgather_small_56PPN} shows the Allgather latency curves for OMB and OMB-Py for larger message sizes with 56 processes per node (full subscription) on 16 nodes. OMB-Py has an overhead that seems to increase with the message size. The overhead starts with 8 microseconds for message size 1B and goes up to 345 microseconds for message size 8KB.
Figure~\ref{fig:allgather_large_56PPN} shows numbers for the same benchmark but for larger message sizes. The overhead for OMB-Py goes up to 41 milliseconds for message size 32KB and it is 16 milliseconds on average for this message size range.

\subsection{Point-to-Point Evaluation on GPU}
\label{sec:evaluation:pt2pt_gpu}
\begin{figure}[!htbp]
    \centering
    \includegraphics[width=0.75\columnwidth]{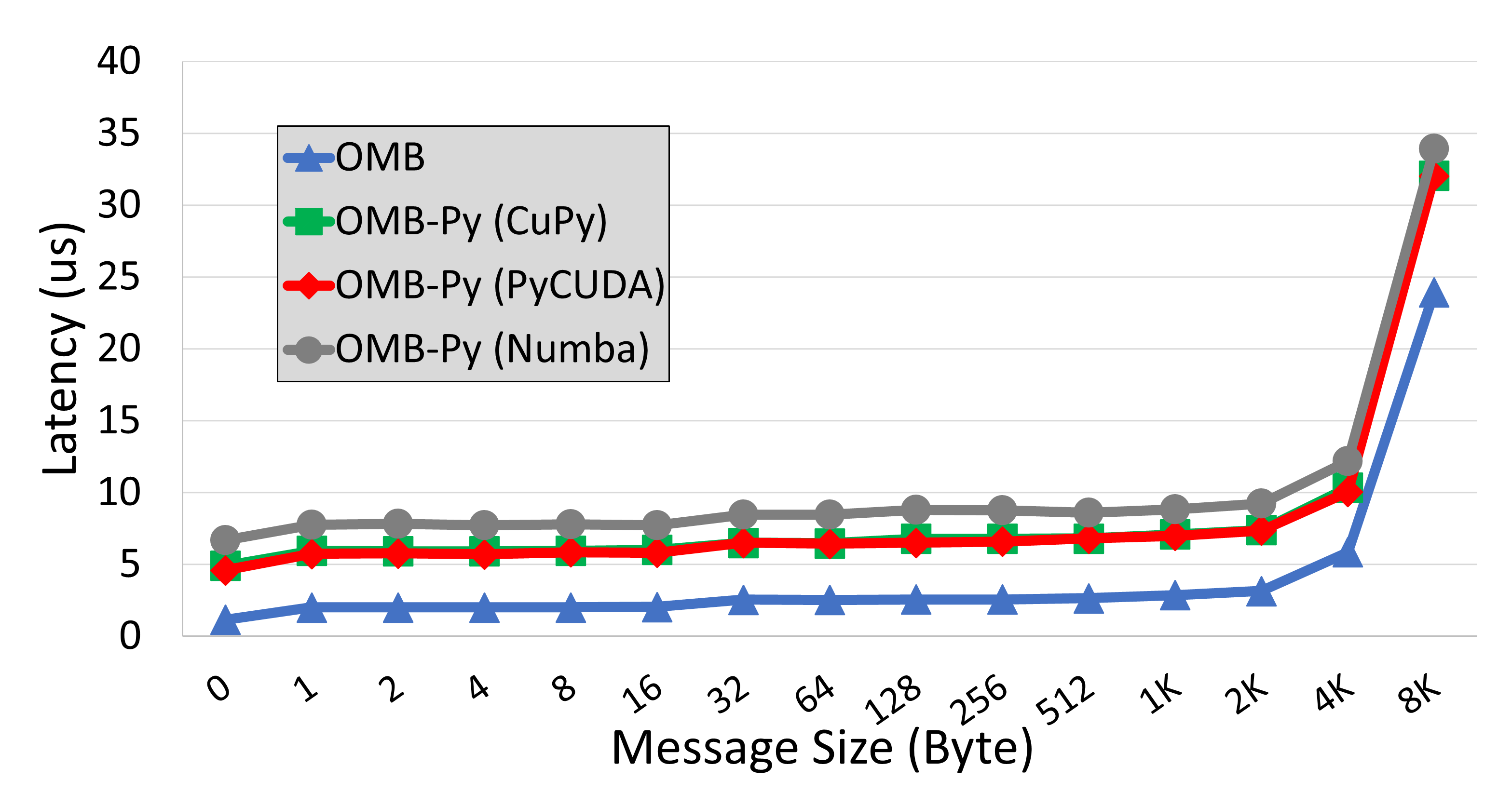}
    \Mycaption{Latency on GPU for small message sizes comparing OMB-Py with different data buffers and OMB on the Bridges-2 cluster.}
    \label{fig:latency_gpu_small}
    \vspace{-1ex}
\end{figure}

\begin{figure}[!htbp]
    \centering
    \includegraphics[width=0.75\columnwidth]{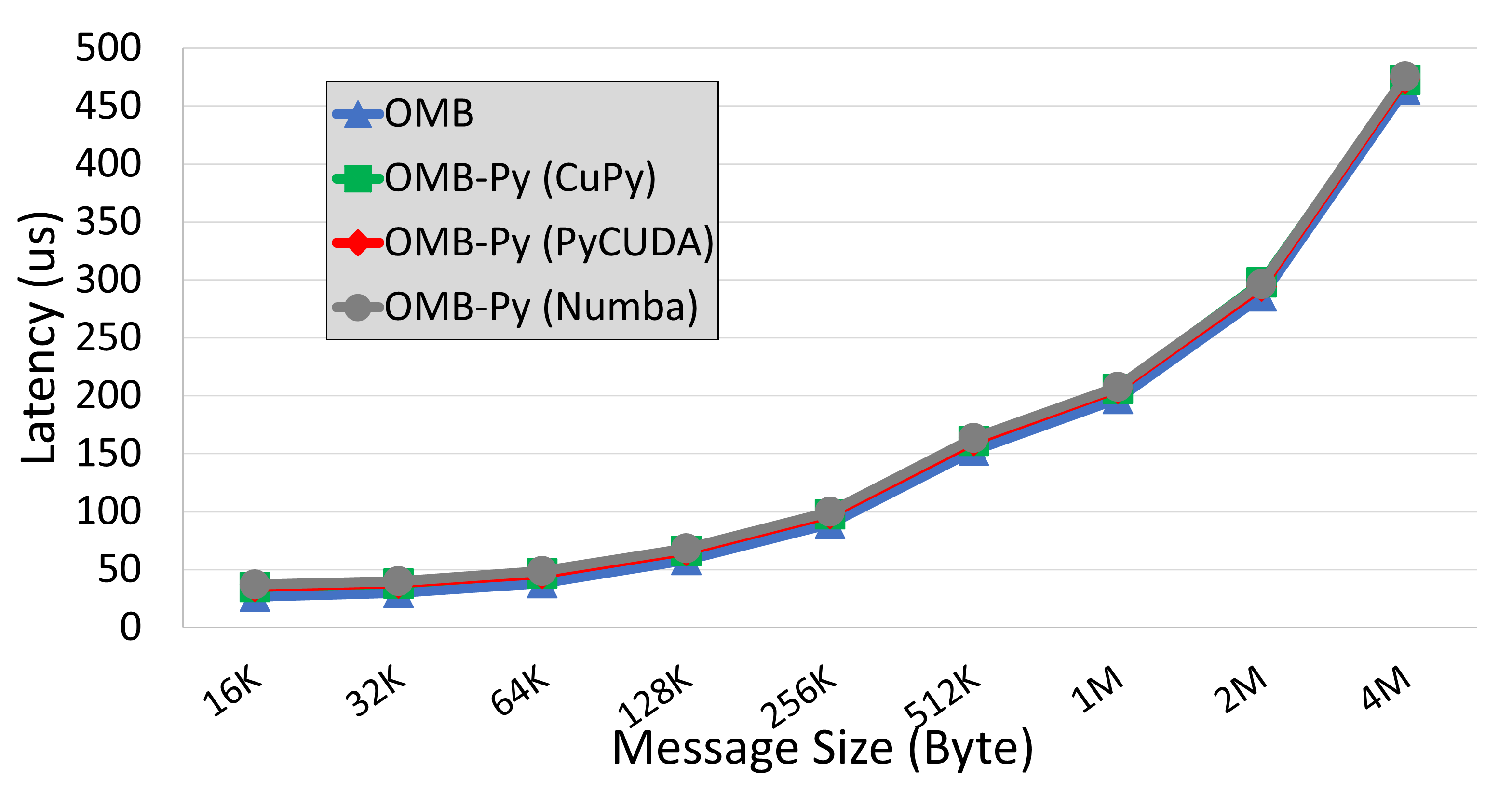}
    \Mycaption{Latency on GPU for small message sizes comparing OMB-Py with different data buffers and OMB on the Bridges-2 cluster.}
    \label{fig:latency_gpu_large}
    \vspace{-0.5ex}
\end{figure}

This subsection provides point-to-point GPU latency evaluation on the Bridges-2 cluster of OMB-Py with three types of GPU-aware data buffers and uses OMB benchmarks as baseline. The three GPU-aware data buffers are: 1) CuPy~\cite{cupy}, 2) PyCUDA~\cite{pycuda}, and 3) Numba~\cite{numba}.
Figure~\ref{fig:latency_gpu_small} shows small message sizes latency curves for OMB and the three data buffers supported by OMB-Py. All OMB-Py numbers have an overhead compared to OMB. CuPy and PyCUDA have very similar numbers and overall perform better than Numba. The average overheads are 4.33, 4.19, and 6.19 microseconds over the OMB numbers for CuPy, PyCUDA, and Numba respectively. 
Figure~\ref{fig:latency_gpu_large} shows latency numbers for the same benchmark but with larger message sizes. Although the four curves look almost identical, there is an average overhead of 8.67, 8.40, and 10.53 microseconds over the OMB numbers for CuPy, PyCUDA, and Numba respectively.

\subsection{Collective Communication Evaluation on GPU}
\label{sec:evaluation:col_gpu}
This subsection provides a sample evaluation of collective MPI communication on GPU the Bridges-2 cluster of OMB-Py with three types of GPU-aware data buffers and using OMB benchmarks as baseline. We chose the Allreduce and Allgather operations to carry out this evaluation. This evaluation is performed on 16 GPUs (2 nodes - 8 GPUs per node).
\subsubsection{Allreduce Evaluation}
\begin{figure}[!htbp]
    \centering
    \includegraphics[width=0.75\columnwidth]{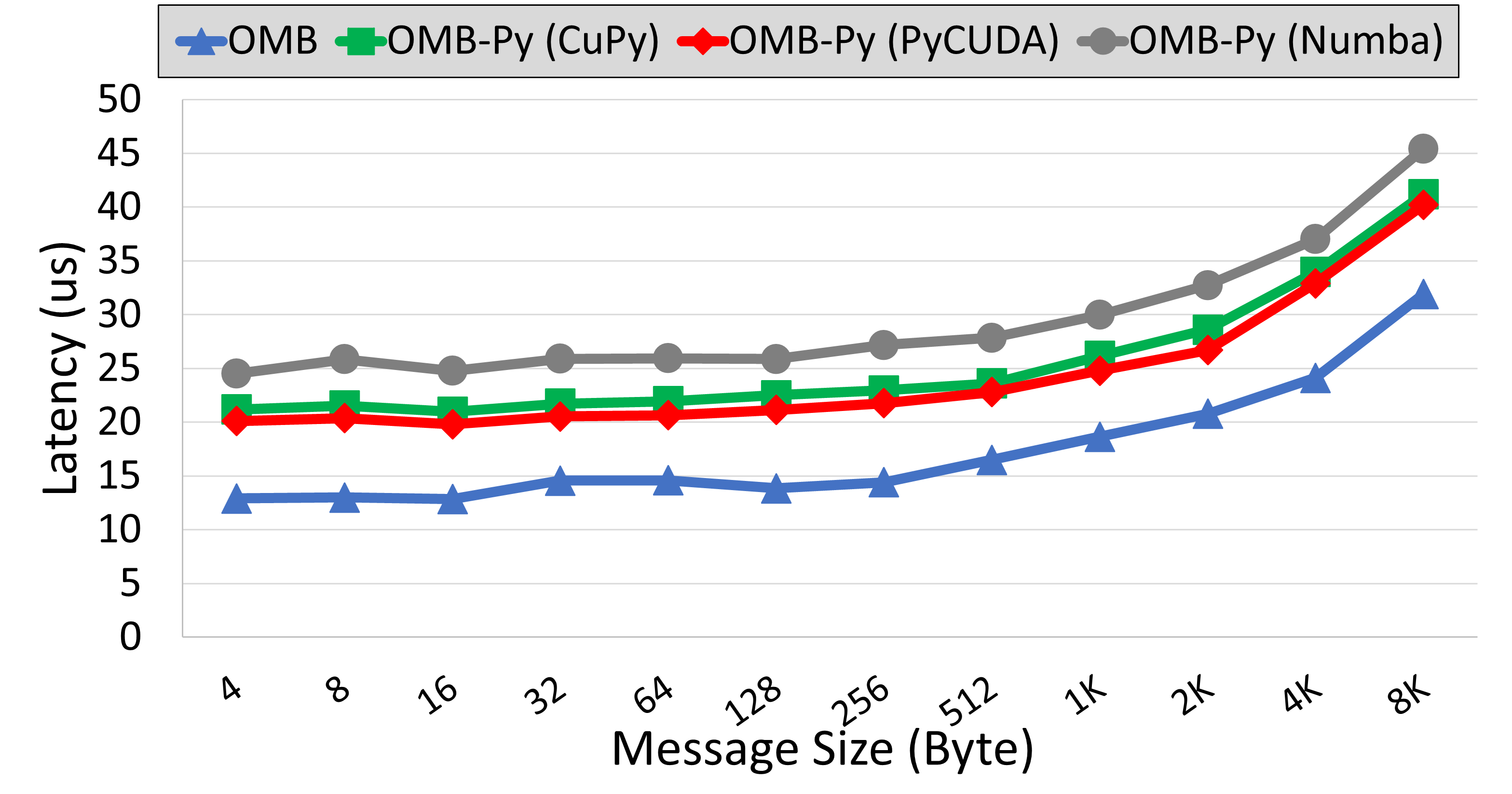}
    \Mycaption{Allreduce GPU latency for small message sizes comparing OMB-Py with different buffers and OMB benchmarks on 16 GPUs (2 nodes - 8 GPUs per node) on the Bridges-2 cluster.}
    \label{fig:gpu_allreduce_small}
    \vspace{-0.5ex}
\end{figure}

Figure~\ref{fig:gpu_allreduce_small} shows small message sizes Allreduce latency curves for OMB and the three data buffers supported by OMB-Py. All OMB-Py numbers have an overhead compared to OMB. CuPy and PyCUDA have very similar numbers and overall perform better than Numba. The average overheads are 8.19, 6.98, and 12.07 microseconds over the OMB numbers for CuPy, PyCUDA, and Numba, respectively. 
Figure~\ref{fig:latency_gpu_large} shows Allreduce latency numbers but with larger message sizes. There is an average overhead of 11.42, 12.17, and 14.76 microseconds over the OMB numbers for CuPy, PyCUDA, and Numba respectively.

\begin{figure}[!htbp]
    \centering
    \includegraphics[width=0.75\columnwidth]{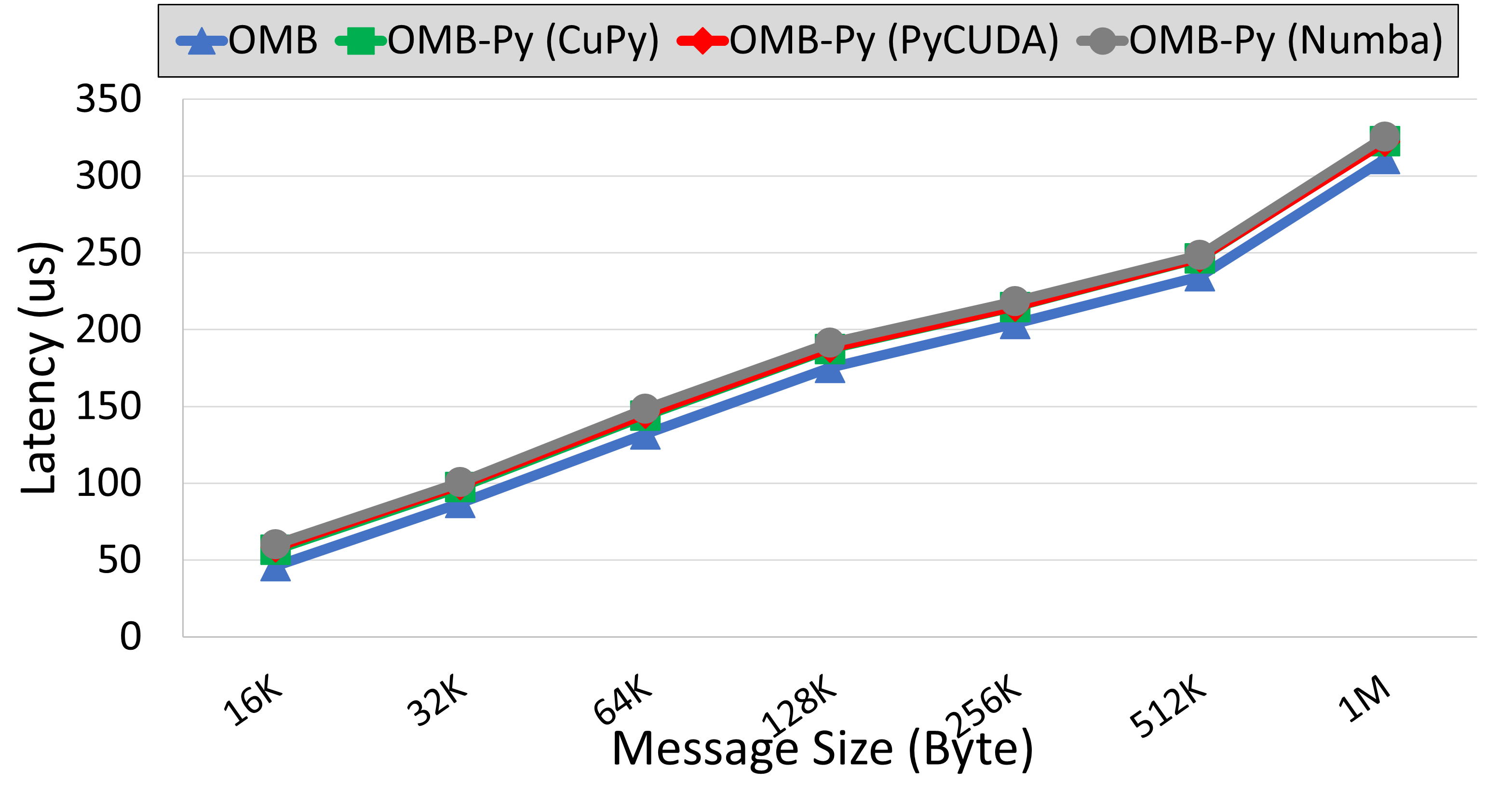}
    \Mycaption{Allreduce GPU latency for large message sizes comparing OMB-Py with different buffers and OMB benchmarks on 16 GPUs (2 nodes - 8 GPUs per node) on the Bridges-2 cluster.}
    \label{fig:gpu_allreduce_large}
    \vspace{-0.5ex}
\end{figure}

\subsubsection{Allgather Evaluation}
\begin{figure}[!htbp]
    \centering
    \includegraphics[width=0.75\columnwidth]{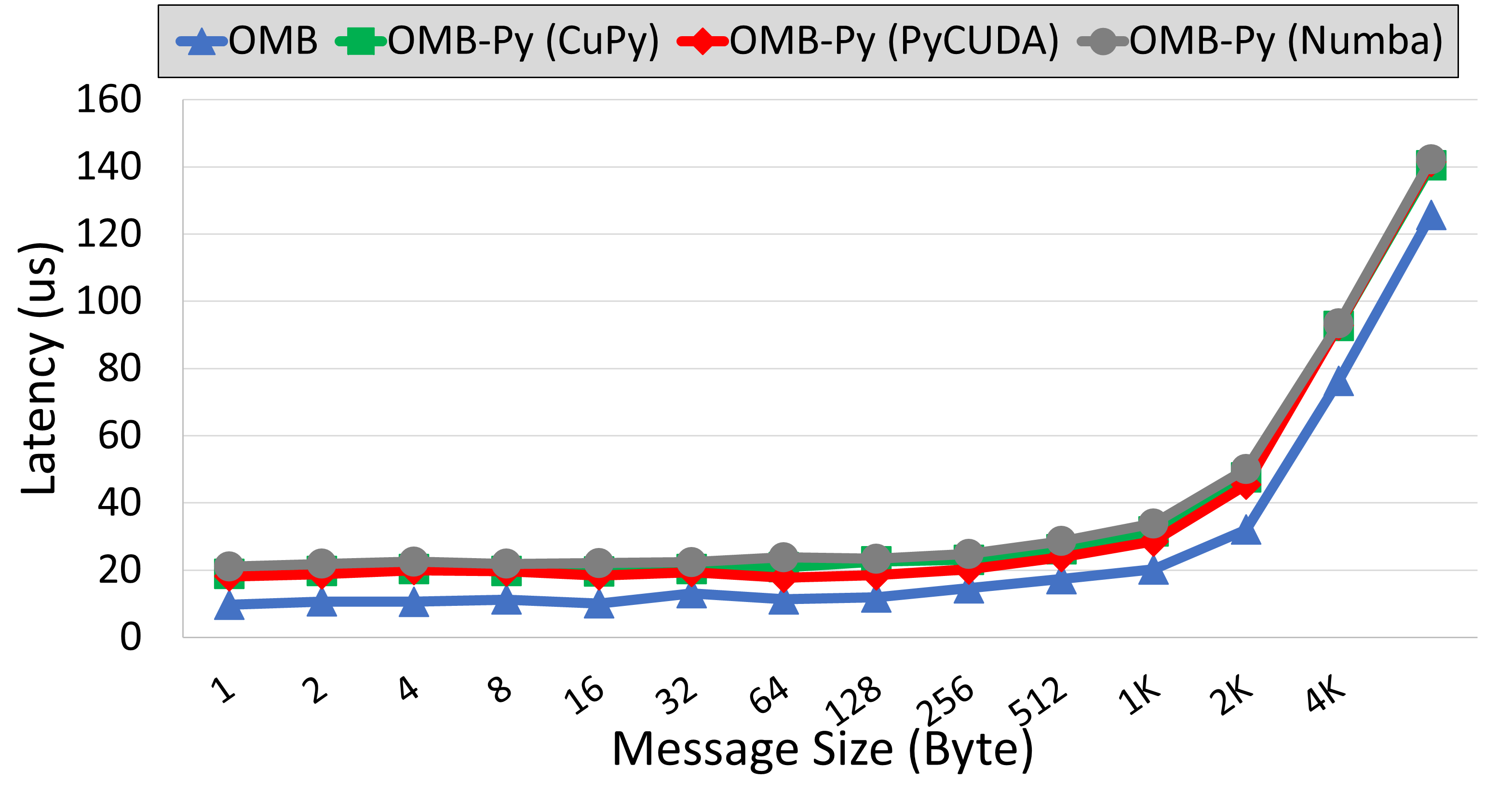}
    \Mycaption{Allgather GPU latency for small message sizes comparing OMB-Py with different buffers and OMB benchmarks on 16 GPUs (2 nodes - 8 GPUs per node) on the Bridges-2 cluster.}
    \label{fig:gpu_allgather_small}
    \vspace{-1ex}
\end{figure}

Figure~\ref{fig:gpu_allgather_small} shows small message sizes Allgather latency curves for OMB and the three data buffers supported by OMB-Py. The average overheads are 10.63, 12.64, and 9.15 microseconds over the OMB numbers for CuPy, PyCUDA, and Numba respectively. 
Figure~\ref{fig:latency_gpu_large} shows Allgather latency numbers but with larger message size. There is an average overhead of 15.04, 16.99, and 19.36 microseconds over the OMB numbers for CuPy, PyCUDA, and Numba respectively.

\begin{figure}[!htbp]
    \centering
    \includegraphics[width=0.75\columnwidth]{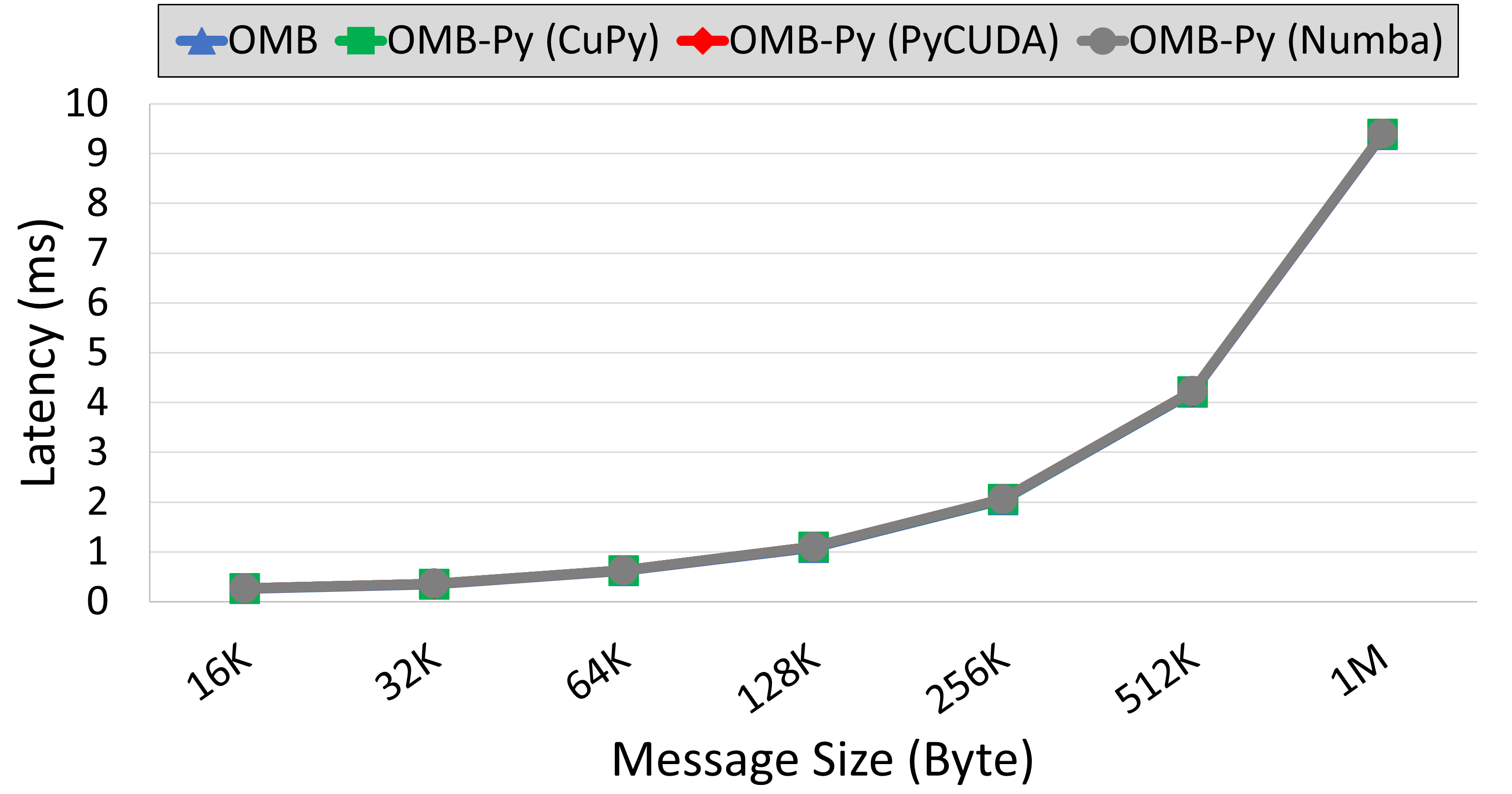}
    \Mycaption{Allgather GPU latency for large message sizes comparing OMB-Py with different buffers and OMB benchmarks on 16 GPUs (2 nodes - 8 GPUs per node) on the Bridges-2 cluster.}
    \label{fig:gpu_allgather_large}
    \vspace{-0.5ex}
\end{figure}

\subsection{OMB-Py Generality}
\begin{figure}[!htbp]
    \centering
    \includegraphics[width=0.75\columnwidth]{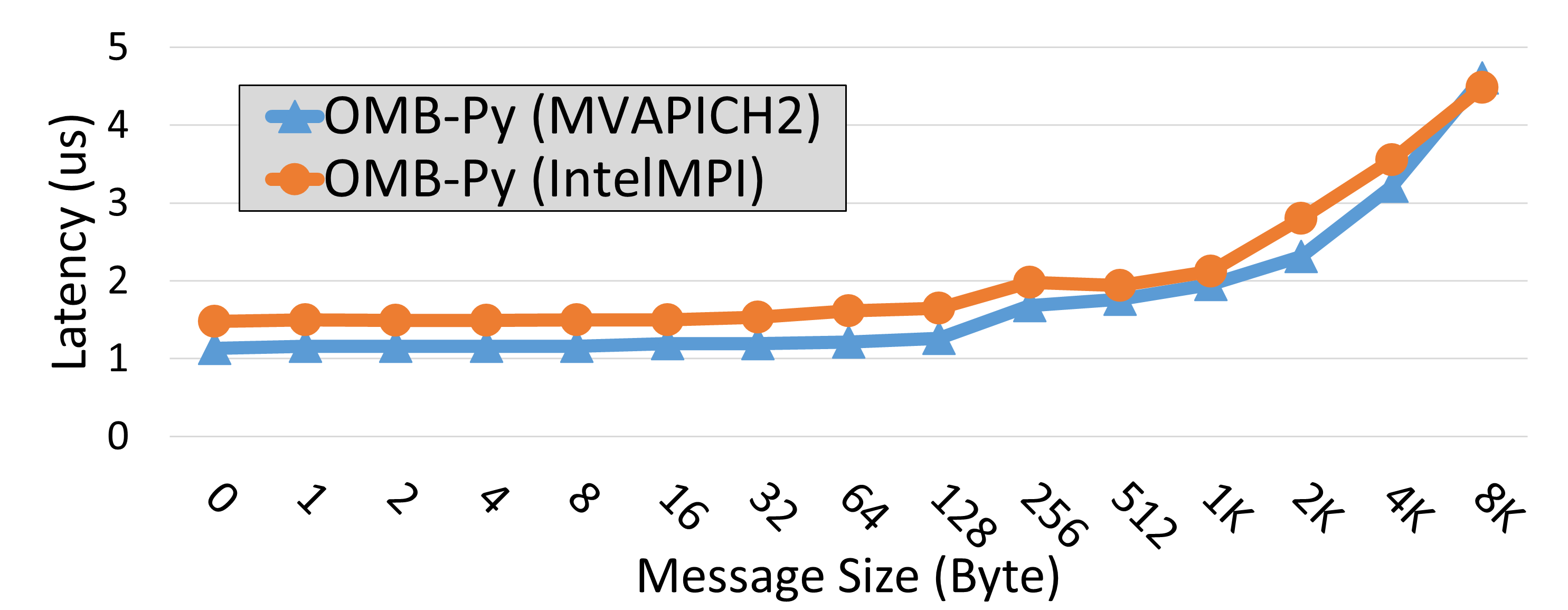}
    \Mycaption{Inter-node CPU latency for small message sizes using OMB-Py with MVAPICH2 and Intel MPI on Frontera.}
    \label{fig:MPIs_latency_small}
    \vspace{-1ex}
\end{figure}

\begin{figure}[!htbp]
    \centering
    \includegraphics[width=0.75\columnwidth]{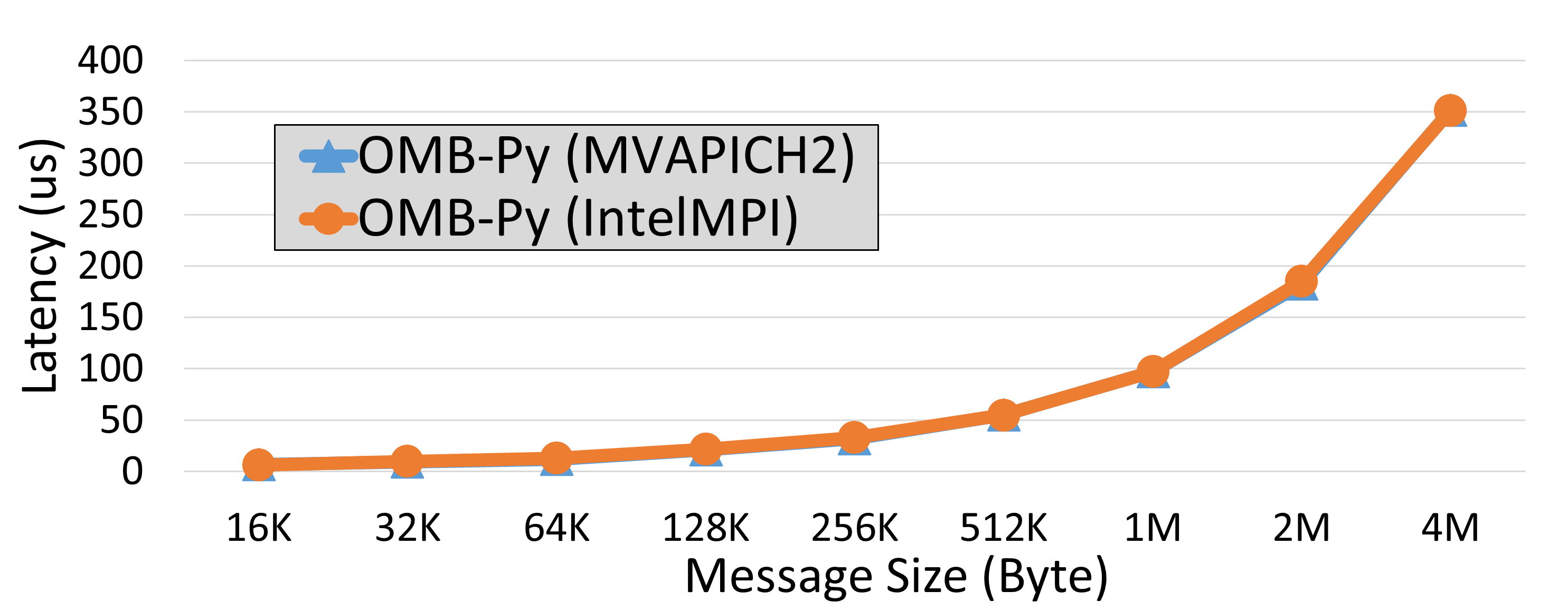}
    \Mycaption{Inter-node CPU latency for large message sizes using OMB-Py with MVAPICH2 and Intel MPI on Frontera.}
    \label{fig:MPIs_latency_large}
    \vspace{-1ex}
\end{figure}

\begin{figure}[!htbp]
    \centering
    \includegraphics[width=0.75\columnwidth]{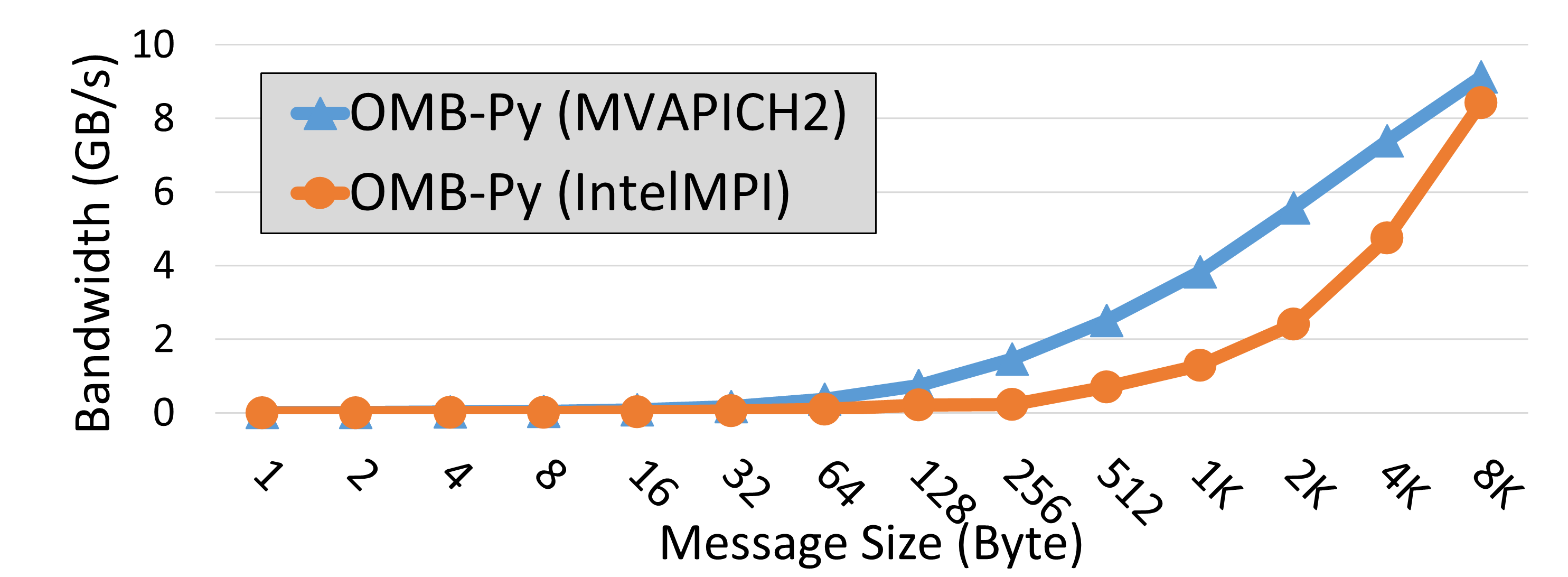}
    \Mycaption{Inter-node CPU bandwidth for small message sizes using OMB-Py with MVAPICH2 and Intel MPI on Frontera.}
    \label{fig:MPIs_bw_small}
    \vspace{-1ex}
\end{figure}

\begin{figure}[!htbp]
    \centering
    \includegraphics[width=0.75\columnwidth]{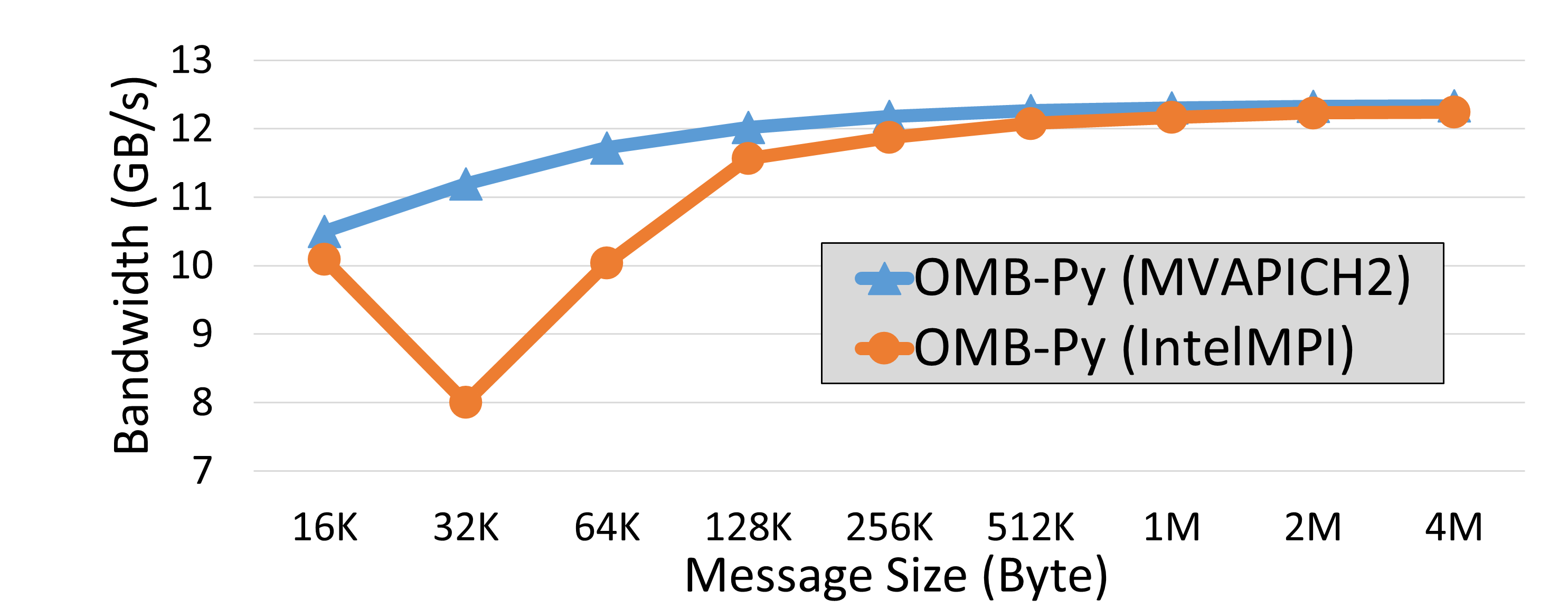}
    \Mycaption{Inter-node CPU bandwidth for large message sizes using OMB-Py with MVAPICH2 and Intel MPI on Frontera.}
    \label{fig:MPIs_bw_large}
    \vspace{-1ex}
\end{figure}

In this section, we show numbers using two different MPI libraries to demonstrate the generality of OMB-Py with regard to MPI implementations. The two libraries we use for this experiment are 1) MVAPICH2 2.3.6~\cite{mvapich2} and 2) Intel MPI Library 19.0.9~\cite{intelmpi}. The following CPU inter-node latency and bandwidth tests are performed on the Frontera cluster. Figure~\ref{fig:MPIs_latency_small} and~\ref{fig:MPIs_latency_large} show numbers for OMB-Py using MVAPICH2 and Intel MPI for small and large message sizes. There is a small difference in latency of 0.36 microseconds on average for all message sizes. Figure~\ref{fig:MPIs_bw_small} and~\ref{fig:MPIs_bw_large} show the bandwidth numbers for both implementations. The average difference is 856 MB/s for all message sizes.


\subsection{Pickle Method Evaluation}
In this section, we conduct latency and bandwidth evaluation for inter-node point-to-point communication on the Frontera cluster using the mpi4py pickle method to serialize the communicated objects. We compare against the direct buffer method in mpi4py.

\subsubsection{Latency}
\begin{figure}[!htbp]
    \centering
    \includegraphics[width=0.75\columnwidth]{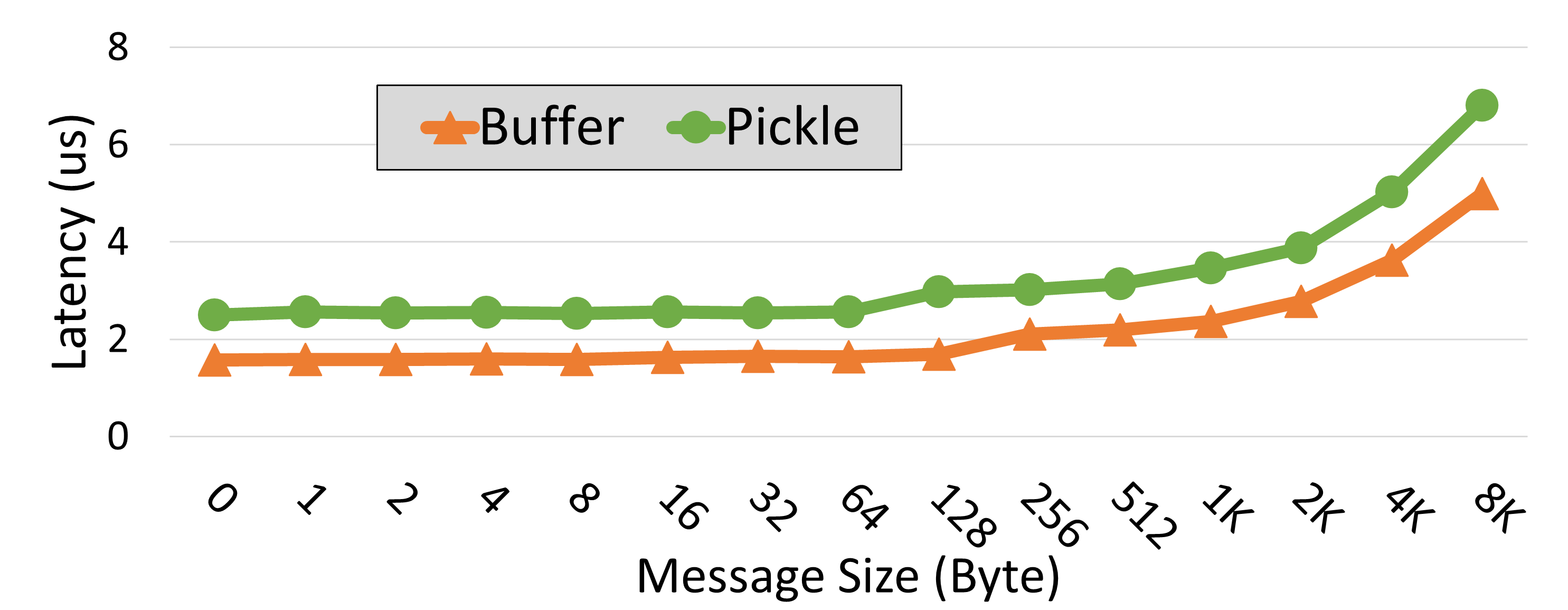}
    \Mycaption{Inter-node CPU latency for small message sizes using OMB-Py to compare the pickle method and direct buffer on the Frontera cluster.}
    \label{fig:inter_pickle_latency_frontera_small}
    \vspace{-1ex}
\end{figure}

\begin{figure}[!htbp]
    \centering
    \includegraphics[width=0.75\columnwidth]{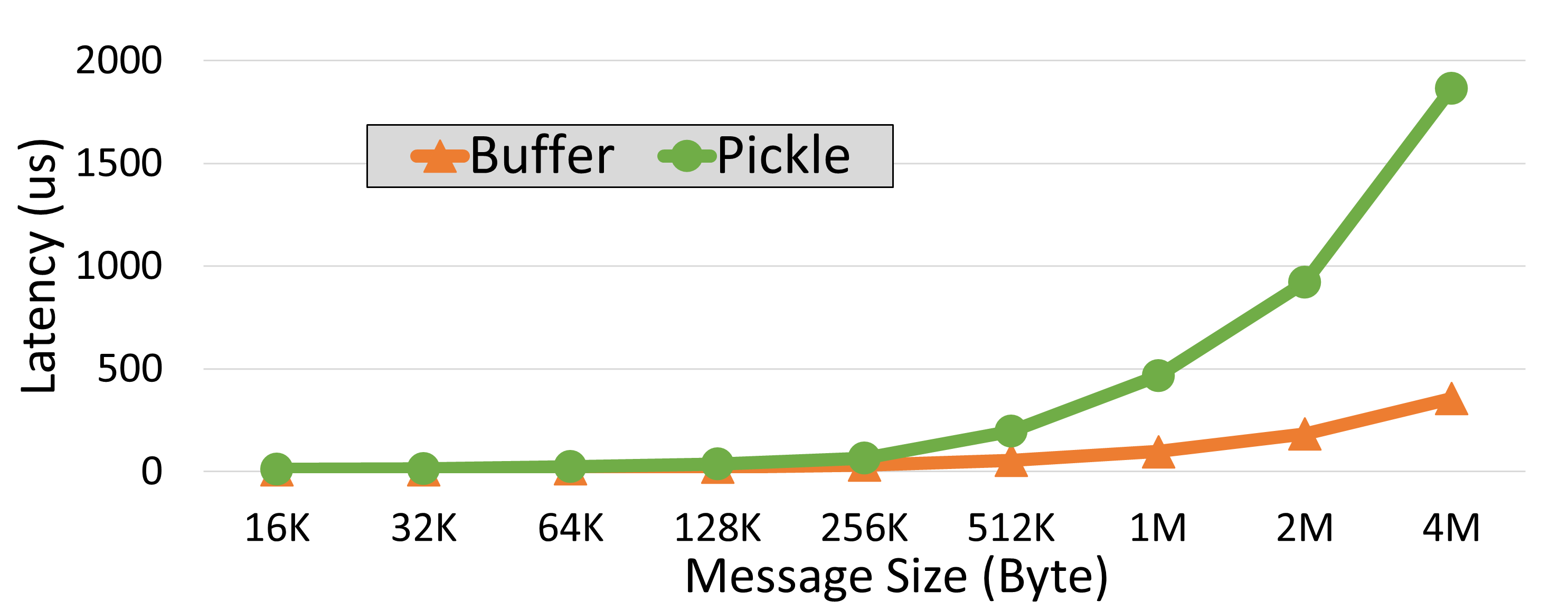}
    \Mycaption{Inter-node CPU latency for large message sizes using OMB-Py to compare the pickle method and direct buffer on the Frontera cluster.}
    \label{fig:inter_pickle_latency_frontera_large}
    \vspace{0.5ex}
\end{figure}

Figure~\ref{fig:inter_pickle_latency_frontera_small} shows the inter-node latency curves using the pickle method and direct buffers method for small message sizes on Frontera. Both latency curves follow the same trend; however, the pickle method has an average overhead of 1.07 microseconds compared to the direct buffer numbers.
Figure~\ref{fig:inter_pickle_latency_frontera_large} shows latency numbers for the same benchmark but for larger message sizes. The two curves start to diverge after message size 64KB with an increasing overhead up to 1,510 microseconds for the pickle method.

\subsubsection{Bandwidth}
\begin{figure}[!htbp]
    \centering
    \includegraphics[width=0.75\columnwidth]{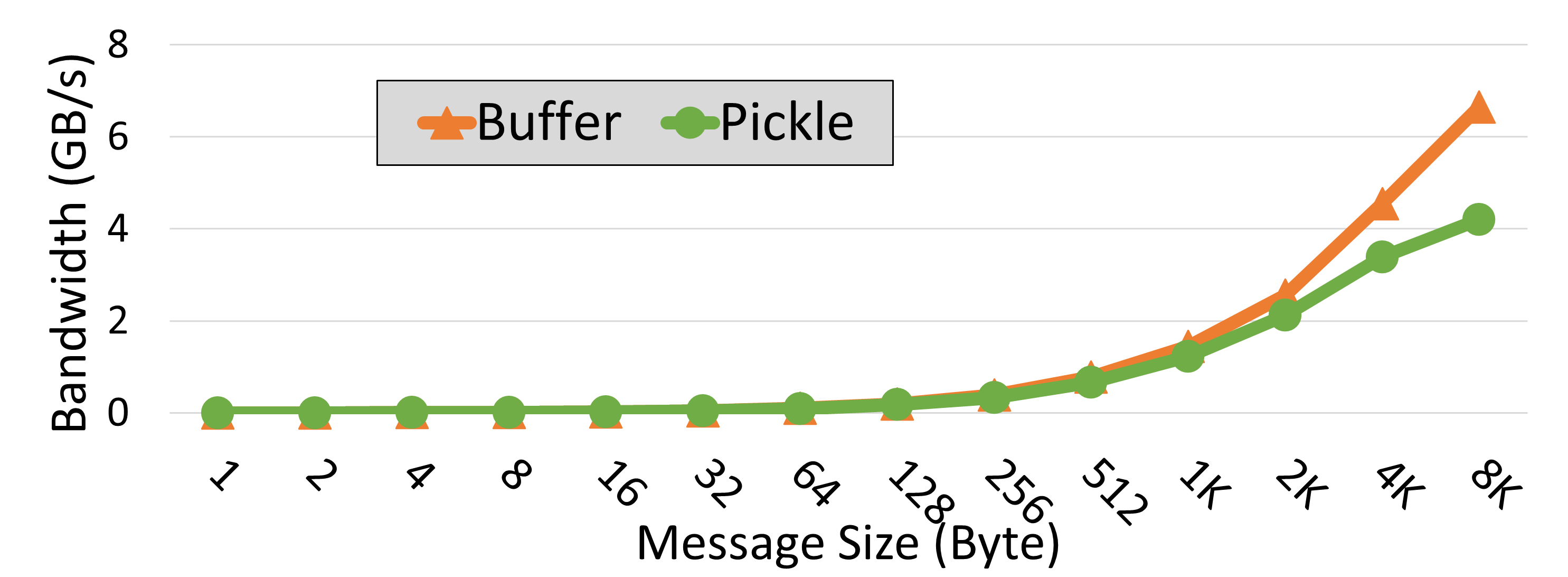}
    \Mycaption{Inter-node CPU bandwidth for small message sizes using OMB-Py to compare the pickle method and direct buffer on the Frontera cluster.}
    \label{fig:inter_pickle_bw_frontera_small}
    \vspace{-0.5ex}
\end{figure}

\begin{figure}[!htbp]
    \centering
    \includegraphics[width=0.75\columnwidth]{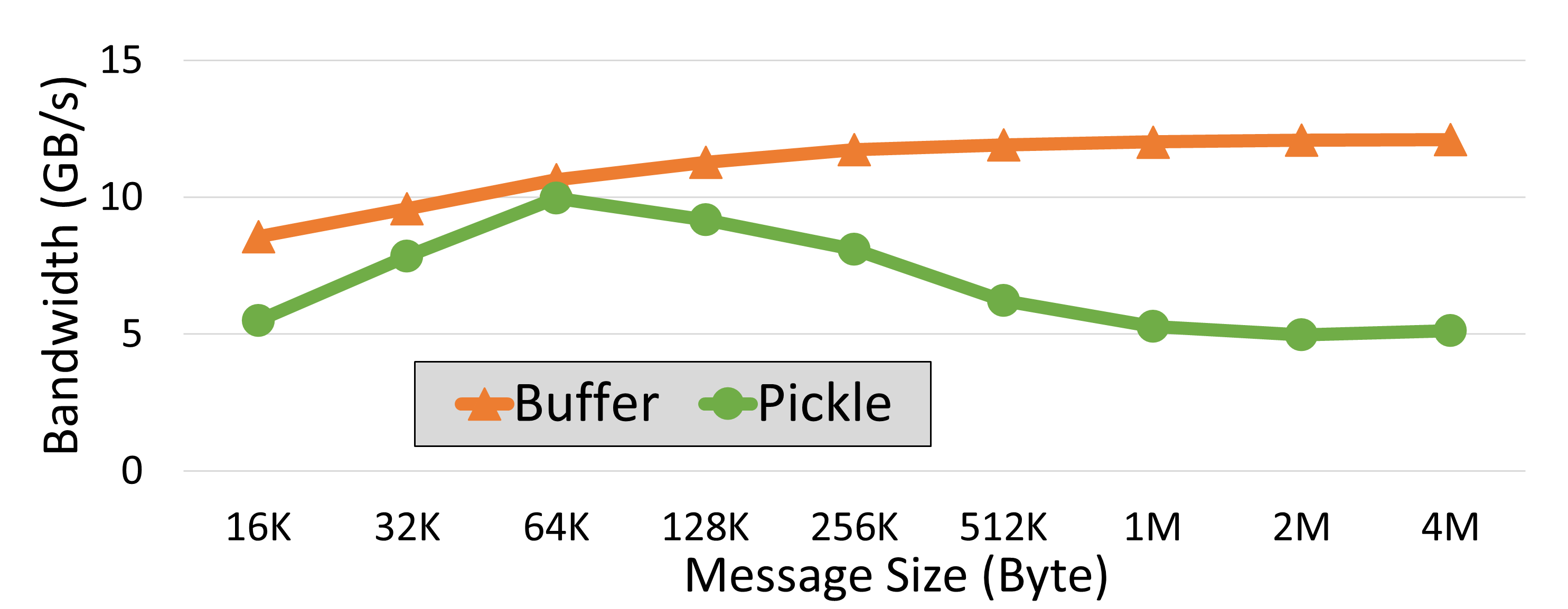}
    \Mycaption{Inter-node CPU bandwidth for large message sizes using OMB-Py to compare the pickle method and direct buffer on the Frontera cluster.}
    \label{fig:inter_pickle_bw_frontera_large}
    \vspace{-0.5ex}
\end{figure}

Figure~\ref{fig:inter_pickle_bw_frontera_small} shows the inter-node bandwidth curves in GB/s using the pickle method and the direct buffers for small message sizes on Frontera. Bandwidth for small message sizes (up to 1KB) looks similar; however, the pickle method starts to have an increasing overhead up to 2.4GB/s for message size 8KB.
Figure~\ref{fig:inter_pickle_bw_frontera_large} shows bandwidth numbers for the same benchmark but for larger message sizes. For this message size range the pickle method starts to catch up but drops in performance again after message size 64KB.
\section{Discussion and Summary of Results}
The evaluation shows an overall small overhead in latency for MPI operations in Python compared to C. On the four clusters we used for evaluation, we notice the same trend for the OMB-Py and OMB numbers. The overhead across the small message size range is almost constant. Table~\ref{table:overhead} shows the average CPU overhead for the latency benchmark (intra-node and inter-node), and the Allreduce benchmark on 16 nodes on the Frontera cluster. Even though the overhead increases with message size, it is more noticeable in small message sizes since the latency is already small. The overhead is relatively negligible for larger messages.

Table~\ref{table:overhead} also shows the average overhead using three different buffers (CuPy, PyCUDA, and Numba) on the Bridges-2 cluster. CuPy and PyCUDA give the best MPI communication performance compared to Numba which shows more overhead in both point-to-point and collective operations as shown in Figures~\ref{fig:latency_gpu_small}, ~\ref{fig:latency_gpu_large}, ~\ref{fig:gpu_allreduce_small}, and~\ref{fig:gpu_allreduce_large}. The difference in overhead depends on how each of these three implementations copy data from/to the GPU.
In order to determine the source of overhead caused by the Python/Cython layer over the native MPI libraries, we perform comprehensive profiling of the mpi4py Allreduce function. The following analysis is done on the Bridges-2 cluster using 16 GPUs (2 nodes - 8 GPUs per node) using three types of GPU buffers (CuPy, PyCUDA, and Numba). The mpi4py Allreduce function can be characterized as consisting of two phases: 1) a staging phase to perform checks and links of the Python send and receive buffers in Cython, 2) an execution phase which mainly calls the implementation of the MPI operation provided by the underlying MPI library. In the staging phase, arguments including count and type of communicated objects are forwarded to two functions (cro\_send and cro\_recv) in order to get pointers to the send and receive buffers. These two functions account for the majority of the overhead of the Cython code. 80\% to 90\% of the overall overhead is spent on preparing the send and receive buffers.
\begin{figure*}[!htbp]
    \centering
    \includegraphics[width=0.93\textwidth]{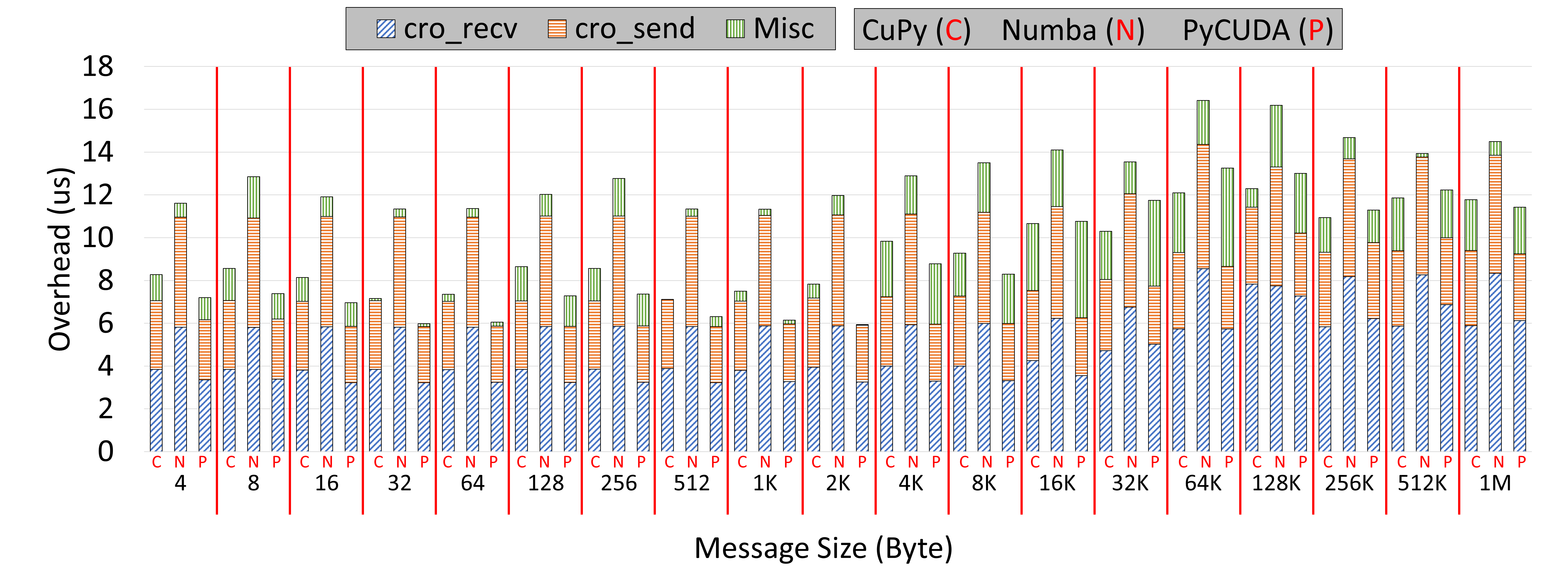}
    \Mycaption{Allreduce GPU overhead analysis using CuPy, Numba, and PyCUDA buffers on 16 GPUs (2 nodes - 8 GPUs per node) on the Bridges-2 cluster.}
    \label{fig:overhead_gpu_allreduce}
    \vspace{0.5ex}
\end{figure*}
Figure~\ref{fig:overhead_gpu_allreduce} shows the different sources that are contributing to the overhead for different GPU buffers and message sizes. We observe the following for each buffer type: 
\begin{itemize}
\item CuPy: The average total overhead is 9.80 microseconds. Preparing the receive buffer contributes to 49\% on average across the different message sizes of the overall overhead. 35\% is spent on preparing the send buffer and 16\% is spent on error checks and other miscellaneous procedures performed in the Cython layer. 
\item Numba: The average total overhead is 13.42 microseconds. Preparing the receive buffer contributes to 50\% on average across the different message sizes of the overall overhead. 40\% is spent on preparing the send buffer and 10\% is spent on error checks and other miscellaneous procedures. 
\item PyCUDA: The average total overhead is 9.57 microseconds. Preparing the receive buffer contributes to 48\% on average across the different message sizes of the overall overhead. 32\% is spent on preparing the send buffer and 20\% is spent on error checks and other miscellaneous procedures.
\end{itemize}
Using the mpi4py pickle method shows an overhead of 1.07 microseconds for small message sizes and 1,510 microseconds for large message sizes on average compared to using direct buffers in the inter-node latency benchmark. This is the expected behavior as serializing larger messages takes longer time. Finally, we demonstrate the ability of OMB-Py to support different MPI implementations by presenting numbers for both the MVAPICH2 and Intel MPI libraries. OMB-Py numbers collected with Intel MPI show larger latency of 0.36 microseconds on average for all message sizes compared to numbers collected with MVAPICH2.

\begin{table}
\caption{Average overhead using OMB-Py compared to OMB on CPU and GPU (three different buffers) using latency and allreduce benchmarks.}
\vspace{-1.0ex}
\centering
\arrayrulecolor{black}
\begin{tabular}{|c|c|l|l|c|c|c|} 
\hline
\rowcolor[rgb]{1,0.82,0.49} {\cellcolor[rgb]{1,0.82,0.49}}                                                                                                       & \multicolumn{3}{c|}{Overhead (CPU) [us]}                            & \multicolumn{3}{c|}{Overhead (GPU) [us]}    \\ 
\hhline{|>{\arrayrulecolor[rgb]{1,0.82,0.49}}->{\arrayrulecolor{black}}------|}
\rowcolor[rgb]{1,0.82,0.49} \multirow{-2}{*}{{\cellcolor[rgb]{1,0.82,0.49}}\begin{tabular}[c]{@{}>{\cellcolor[rgb]{1,0.82,0.49}}c@{}}Message\\Size\end{tabular}} & \multicolumn{1}{l|}{Intra} & \multicolumn{1}{c|}{Inter} & Allreduce & CuPy & PyCUDA & \multicolumn{1}{l|}{Numba}  \\ 
\hline
Small                                                                                                                                                            & 0.44                       & 0.43                       & 0.93      & 4.33 & 4.19   & 6.19                        \\ 
\hline
Large                                                                                                                                                            & 2.31                       & 0.63                       & 14.13     & 8.67 & 8.40   & 10.53                       \\
\hline
\end{tabular}
\label{table:overhead}
\vspace{-3.0ex}
\end{table}
\section{Related Work}
\label{sec:related_work}
In~\cite{OMB-GPU} Bureddy et al. develop OMB-GPU which is an extension to the OMB library that enables the comparison of MPI implementations on GPU clusters. They conduct a comprehensive evaluation for GPU MPI communication performance using several point-to-point and collective benchmarks. However, Python support is not provided with this package. In~\cite{par_python},~\cite{ mpi4py_mpi2}, Dalcin et al. conduct performance evaluation for MPI in Python using mpi4py and NumPy arrays as data buffers. They report latency, bandwidth, and alltoall numbers on CPUs comparing MPI in C to Python using direct buffers and the pickle method. However, at the time when those papers were released, the MPI-2 standard was still mainly used and there was no support for GPU-aware MPI communication in Python. In~\cite{perf_mpi4py}, Smith uses mpi4py and NumPy to perform several tests to measure the performance of distributed algorithms like Graph 500 in Python on HPC systems. In~\cite{wazir2019performance} Wazir et all. implement using Python and mpi4py a number of distributed algorithms like the Monte Carlo’s method and prime number generator and measure the gain in performance on a Raspberry PI cluster compared to sequential execution.
\section{Conclusion}
\label{sec:conclusion}
In this paper, we presented OMB-Py which is a micro-benchmarks package that offers point-to-point and blocking collective tests to evaluate the performance of MPI implementations on HPC systems using Python and mpi4py for MPI-Python bindings. OMB-Py supports benchmarking for both CPUs and GPUs as communication devices with a wide range of user flags to run customizable tests. The proposed design is evaluated against OMB as a baseline to characterize the performance of MPI operations with Python. We conduct our experiments on four HPC systems and we use three GPU-aware data buffers (CuPy, PyCUDA, and Numba) in the GPU evaluation. Additionally, we evaluate the performance of the pickle methods in mpi4py compared to using direct buffers. We gain the following insight by performing evaluation using the proposed design: 1) Small overhead in latency for MPI operations in Python compared to C, which is more noticeable in small message sizes compared to large message sizes, 2)
similar performance trends for MPI operations in Python on the 3 CPU architectures we evaluated, 3) CuPy and PyCUDA as GPU-aware Python data buffers give better MPI communication performance compared to Numba, 4) 80\%-90\% of the overhead of mpi4py over native MPI libraries comes from preparing the send and receive buffers to link the Python objects in the Cython layer.
Finally, we plan to publicly release OMB-Py to benefit the Python and HPC community. To the best of our knowledge, this is the first comprehensive MPI micro-benchmarks package that supports Python.

%\section*{Acknowledgment}

\bibliographystyle{ieeetr}

\end{document}